\documentclass[journal,twoside,web]{IEEEtran}

\usepackage{amsmath,amssymb,amsfonts}
\usepackage{epsfig}
\usepackage{cite}
\usepackage{amsfonts}
\usepackage{amssymb}
\usepackage{mathtools}
\usepackage{booktabs}
\usepackage{enumerate}
\usepackage{epstopdf}
\usepackage{savesym}
\usepackage{algorithm}
\usepackage{algorithmic}
\usepackage{cases}
\usepackage{stix}%

\newtheorem{theorem}{Theorem}
\newtheorem{corollary}{Corollary}
\newtheorem{definition}{Definition}
\newtheorem{lemma}{Lemma}

\newtheorem{assumption}{Assumption}
\newtheorem{remark}{Remark}
\newtheorem{example}{Example}

\newcommand{\zeb}{\mathbf{0}}

\newcommand{\vb}{\mathbf{v}}
\newcommand{\wb}{\mathbf{w}}
\newcommand{\xb}{\mathbf{x}}
\newcommand{\yb}{\mathbf{y}}
\newcommand{\zb}{\mathbf{z}}

\newcommand{\eb}{\mathbf{e}}
\newcommand{\rb}{\mathbf{r}}

\newcommand{\Jb}{\mathbf{J}}
\newcommand{\Ab}{\mathbf{A}}
\newcommand{\Bb}{\mathbf{B}}
\newcommand{\Cb}{\mathbf{C}}

\newcommand{\Fb}{\mathbf{F}}
\newcommand{\Lb}{\mathbf{L}}
\newcommand{\Rb}{\mathbf{R}}

\newcommand{\Wb}{\mathbf{W}}
\newcommand{\Kb}{\mathbf{K}}

\newcommand{\Pb}{\mathbf{P}}
\newcommand{\Qb}{\mathbf{Q}}
\newcommand{\Ib}{\mathbf{I}}
\newcommand{\Mb}{\mathbf{M}}

\usepackage{textcomp}
\def\BibTeX{{\rm B\kern-.05em{\sc i\kern-.025em b}\kern-.08em
		T\kern-.1667em\lower.7ex\hbox{E}\kern-.125emX}}

\begin{document}

	\title{Exponential Consensus of Multiple Agents over  Dynamic Network Topology: Controllability, Connectivity, and Compactness}
	\author{Qichao Ma, Jiahu Qin, Brian D. O. Anderson, and Long Wang%
		\thanks{This work was supported in	part by the Naional Natural Science Foundation of China under Grants	61922076,  62003322, and 62036002, in part by Science and Technology Major Project of
			Anhui Province (202203a06020011), and  in part by the
			National Key Research and Development Program of China under Grant
			S02022AAA010129.  (\emph{Corresponding author: Jiahu	Qin.}) }
		\thanks{%
			Q.~Ma is with the Department of Automation, University of Science and Technology of China,
			Hefei 230027, China
			(e-mail: mqc0214@ustc.edu.cn).}
	    \thanks{J. Qin is with the Department of Automation, University of Science and Technology of China, Hefei 230027, China, and also with the Institute of Artificial Intelligence, Hefei Comprehensive National Science Center, Hefei 230088, China (e-mail: jhqin@ustc.edu.cn).}
		\thanks{B. D. O. Anderson is with the School of Engineering, Australian National University, Acton ACT 2601, Australia (e-mail: brian.anderson@anu.edu.au).}
	    \thanks{L.~Wang is with the Center for Systems and Control, College of Engineering, Peking University, Beijing 100871, China (e-mail: longwang@pku.edu.cn).}
	}
	\maketitle

\begin{abstract}  
This paper investigates the problem of securing exponentially fast consensus (exponential consensus for short) for identical agents with finite-dimensional linear system dynamics over dynamic network topology. Our aim is to find the weakest possible conditions that guarantee exponentially fast consensus using a Lyapunov function consisting of a sum of terms of the same functional form. 
We first investigate necessary conditions, starting by examining the system (both agent and network) parameters. It is found that  controllability of the linear agents is necessary for reaching   consensus. Then,  to work out  necessary conditions incorporating the network topology, we  construct a set of Laplacian  matrix-valued functions.  The precompactness of this set of functions is   shown to be a significant generalization of   existing assumptions on network topology, including the common assumption that the edge weights are bounded piecewise constant functions or continuous functions.  With the aid of such a precompactness assumption and restricting the Lyapunov function to one consisting of a sum of terms of the same functional form, we prove that a joint $(\delta, T)$-connectivity condition on the network topology is  necessary for exponential consensus. Finally, we investigate how the above two ``necessities'' work together to guarantee exponential consensus. To partially address this problem, we define a synchronization index to characterize the interplay between agent parameters and network topology. Based on this notion, it is shown   that by designing a proper feedback matrix and under the precompactness assumption, exponential consensus can be reached globally and uniformly if the joint $(\delta,T)$-connectivity and controllability conditions are satisfied, and  the synchronization index is not less than one.  
\end{abstract}


\section{Introduction\label{sec:intro}}

\IEEEPARstart{C}{onsensus} is ubiquitous  in distributed control,  estimation, and computation,  see \cite{Saber} and \cite{Jadbabaie2003}, as representatives of a massive literature. It refers to agreement of a network of individual agents on a quantity of interest, e.g.,  position, opinion, and estimation \cite{Jadbabaie2003,Ye2018TAC}.
Over the last decades,  consensus analysis has attracted significant attentions in systems and control and also in social network analysis \cite{QinTAC2016,WangAUTO2015,SuTAC2016,XiaoTAC,MoreauTAC}. 

Very recently, it has been found applicable to various interesting scenarios, for example shortest path planning \cite{Zhang},   distributed optimization algorithm design \cite{Nedic}, and   resilient  estimation under attack over networks \cite{Chen}. These potential applications, consequently,  motivate  further investigation of coordination/cooperation of networked agents.  With this inspiration, in this paper we focus  on development of the weakest possible conditions to guarantee  exponentially fast consensus;  the exponential property is generally desirable and offers  better performance and more robustness against noise, parameter perturbations, nonlinearity, etc. Such weak conditions also  help us gain insight into how the networked high-order\footnote{In this paper,  ``high-order" means that each agent has high-order   dynamics. Note that to follow convention, the networked linear agents are described by first-order differential equations and their states have a high dimension. } linear agents interact with each other and evolve.

To date, various necessary and/or sufficient connectivity conditions for achieving consensus under dynamic network topology (i.e., the edge weights are functions of time) have been developed for integrator agents \cite{MoreauTAC,CaoMSIAM2008,Shi2013,AndersonTAC2017,XiaoTAC2028,Altafini2013}. Of particular interest is the joint $(\delta,T)$-connectivity condition (refer to Definition \ref{joint-connectivity} in the next section). It is shown in \cite{AndersonTAC2017} to be necessary and sufficient for exponential consensus of continuous-time integrator agents with undirected dynamic  network topology.  
Therefore, it is  of theoretical interest to analyze whether joint $(\delta,T)$-connectivity  is still necessary for exponential consensus of high-order linear agents. This question also motivates the current work.

Achieving consensus for high-order linear systems with self-dynamics over a dynamic network topology has been studied over the past decades. Existing literature usually imposes stringent conditions on system parameters (including system and input matrices) \cite{QinTAC2014,YangTAC2016,LuTAC2017,MengHaofei,Liu,Meng,Abdessameud} or network topology \cite{KimAUTO2013,WangAUTO2015,BackTAC2017,ValcherAUTO2017,QinTNNLS}. A frequent  connectivity  condition or a well-defined averaged connectivity  condition is commonly proposed to accommodate consensus analysis \cite{WenTNNLS2015,KimAUTO2013,WangAUTO2015,BackTAC2017}. For instance,   piecewise continuous networks are assumed to have \textit{uniformly bounded weights} \cite{QinTNNLS}, i.e., the connectivity of the network topology remains unchanged. Other works usually carry out consensus analysis with  a full-rank input matrix \cite{QinTAC2014} or a non-expansive system matrix.    Synchronizing heterogeneous systems  is taken into consideration in \cite{LuTAC2017,MengHaofei,Abdessameud}, and \cite{Liu}. By designing full-state coupled dynamic controllers or reference signals, the synchronization problem is transformed into a consensus problem for integrator agents \cite{LuTAC2017,MengHaofei,Abdessameud,Liu}. Recently, consensus of nonexpansive time-varying finite-dimensional linear agents with a full-rank input matrix  was investigated in  \cite{Meng}.  A common weakness  of the above works \cite{QinTAC2014,YangTAC2016,LuTAC2017,MengHaofei,Liu,Meng,Abdessameud} is that the coupling configurations take full-state coupled forms. This relaxes the assumption on network connectivity and makes the consensus analysis easier.

If the input matrix takes non-full row rank and simultaneously a weak connectivity assumption is imposed, as far as we know, most results  reported for consensus of high-order linear systems are based on having an undirected network topology \cite{SuTAC2012,Wang,Ma-TCYB,Ma-TAC}.  In \cite{SuTAC2012},  \textit{marginally stable} linear systems with a non-full row rank input matrix  are considered. These linear agents communicate over piecewise constant network topology with a positive dwell time (see Section \ref{sec: compact set of functions} for the definition of ``dwell time''). The authors prove that, with a properly chosen feedback matrix, as long as a uniform connectivity condition (a special case of joint $(\delta,T)$-connectivity) and an observability condition  are satisfied, asymptotic consensus can be ensured \cite{SuTAC2012}. Consensus of \textit{neutrally stable} linear systems over either piecewise constant network topology or continuously time-varying  network topology (i.e., edge weights are continuous functions of time) has recently been studied in  \cite{Ma-TAC} and \cite{Ma-TCYB}, separately. A subspace-based method \cite{Ma-TCYB} and a uniform complete observability-based approach \cite{Ma-TAC} are, respectively, developed. With these methods, necessary and sufficient conditions for consensus have been successfully derived \cite{Ma-TAC,Ma-TCYB}.  In \cite{Wang}, it is  assumed that the network topology is piecewise constant with a positive dwell time. The constraint on the system matrix is relaxed    such that the linear agents only need to be controllable \cite{Wang}.  The authors show that consensus can be reached asymptotically with a suitably designed feedback matrix if the Lyapunov exponent  is less than the synchronizability exponent, which is defined  to describe a quantitative characteristic of the network topology \cite{Wang}.

To conclude, although impressive advances have been made  \cite{SuTAC2012,Wang,Ma-TAC,Ma-TCYB} for consensus analysis of finite-dimensional linear systems over an undirected dynamic network topology, there are still several issues left for further consideration. First, consensus conditions imposed on the network topology  or system parameters  are still stringent, e.g., the existence of a positive dwell time  \cite{Wang}. Second, necessary conditions  for convergence to consensus  have rarely been considered.

With the above motivation, we revisit the consensus problem of linear systems under the framework of  undirected networks.  We restrict attention to undirected graphs because (a) how to handle the directed case is at present not clear, and indeed most results related to those of this paper also assume undirected graphs and (b) undirected graphs will be appropriate in many applications (even if in some of these applications, use of directed graphs could also be contemplated if there were supporting theory).

One potential application of our setup is formation control of mobile agents.    Many mobile robots can be described by linear dynamics or are well-linearizable, e.g., the non-holonomic car-like robot operating in the plane \cite{WebbICRA}.   For formation control of these robots in search or rescue missions, the consensus-based formation control algorithm is commonly adopted \cite{MarinaTAC}.
If the robots are equipped with identical wireless communication devices having omnidirectional antennas, then it is reasonable that the communication links are symmetric \cite{HuangWCMC} or at least are approximately symmetric with small perturbations. Normally the longer the communication distance is between two agents, the lower is the signal to noise ratio (since the received signal power decreases with increase in distance \cite{HuangWCMC}). For  communications between distant agents it is reasonable to use a smaller value of $w_{ij}$ than for communications between near agents, thus reflecting the lower reliability of the communication. Hence, the weights of  links are time-varying as the distances between different agents are changing. In the above case, the  network topology can be mathematically characterized by a time-varying undirected weighted graph. Small perturbations of the links may exist, but can be tolerated since we secure exponentially fast consensus.

Our aim is to develop as weak as possible   conditions to guarantee exponential consensus for the linear systems. We consider a very general setup. Specifically, the eigenvalues of the system matrix lie in the \textit{closed right-half plane} and the input matrix is \textit{not} required to be of full row-rank. In addition, the edge weights (of the network topology) are  measurable functions of time and only a mild joint connectivity condition (i.e., joint $(\delta,T)$-connectivity) is imposed.   This setup  includes those studied in \cite{ValcherAUTO2017,SuTAC2012,Ma-TAC,Ma-TCYB} and \cite{Wang} as special cases. However, it poses significant challenges because we need to characterize how the  dynamic network topology, the non-full row rank input matrix, and the unstable system matrix influence agents' evolution simultaneously.  To overcome the challenges, we catch the essence of existing assumptions on ``continuity" of network topology and propose a new \textit{precompact} condition. This condition requires that the edge weights do  not vary too fast and turns out to significantly generalize the conditions on how the network topology varies in existing literature (including but not limited to our own works  \cite{Ma-TAC} and \cite{Ma-TCYB}).  
One advantage of the precompactness condition is that we are able to deal with  piecewise constant and continuous network topologies in the one framework.

The main contributions, which are based on the precompactness condition,  are summarized as follows.
\begin{enumerate}

\item We prove that controllability of the individual finite-dimensional linear systems is necessary for (exponential) consensus   over a dynamic network topology. To the best of our knowledge, this is the first result on necessity of controllability for consensus of linear agents over  a dynamic network topology.

\item With a suitably designed feedback matrix and a time-invariant quadratic Lyapunov function displaying a certain structural constraint, and using the precompactness condition,  we are able to show that a joint $(\delta,T)$-connectivity condition  is necessary  for exponential consensus of linear agents. This generalizes the celebrated result in \cite{AndersonTAC2017} and  is also the first result on necessity of joint $(\delta,T)$-connectivity for exponential consensus of high-order linear agents over dynamic network topology.

\item   We define a new concept in the context of the multiagent problems being treated which we term a \textit{synchronization index}. The novel feature of this index is that it reflects  how system parameters  and network topology  work together to reach consensus. More explicitly, we show that by designing a proper feedback matrix and under the precompactness assumption, exponential consensus can be reached for linear agents globally and uniformly if joint $(\delta,T)$-connectivity and controllability conditions are satisfied, and the synchronization index is not less  than one.
\end{enumerate}

There are some significant differences with existing results. For instance, in \cite{SuTAC2012}, the system matrix is marginally stable and the network topology is piecewise constant. 
Ref. \cite{Wang} allows the system matrix to contain unstable modes but still assumes that the network topology is piecewise constant with a positive dwell time.  Moreover, the authors in \cite{Wang} do not investigate whether their sufficient conditions for consensus are still necessary. Compared to \cite{Ma-TAC,Ma-TCYB}, we are able to deal with piecewise continuous and piecewise constant network topologies in a unified framework with the aid of a new and insightful precompact condition. Moreover, a system matrix may contain unstable modes instead of being neutrally stable, which, roughly speaking, implies more force will be required   to synchronize agents. Hence the techniques to derive both necessary and sufficient conditions are completely different from those (subspace- and uniform-complete-observability-based methods) in  \cite{Ma-TAC,Ma-TCYB}.

The remainder of the paper is arranged as follows. The  problem is formulated in Section \ref{sec:problem formulation}. The precompact set of Laplacian matrix-valued functions on a fixed interval is given in Section \ref{sec: compact set of functions}. Necessary and sufficient conditions for exponential consensus are provided in Section \ref{sec: necessary conditions}. The proofs of the main theorems are given in Section \ref{sec:proofs}, followed by numerical examples in Section \ref{sec:examples}. Finally, the paper is concluded in Section \ref{sec:conclusion}. 

\smallskip

\textit{Notations}: $\mathbb{Z}_+$ represents the set of positive integers.	$\|\cdot\|$ denotes the Euclidean norm of a real vector  or spectral norm of a real matrix. $\Ib_s$ is the identity matrix with dimension $s\in\mathbb{Z}_+$. $\mathbf{1}_N=[1,\ldots,1]^\top\in\mathbb{R}^N$. 	$\mathrm{diag}\{\Pi_1,\ldots,\Pi_N \}$ denotes a diagonal matrix whose $i$-th diagonal entry is $\Pi_i$. Given a symmetric real matrix $\Mb\in\mathbb{R}^{n\times n}$, let  $\lambda_{\min}(\Mb)=\lambda_1(\Mb)\leq\cdots\leq\lambda_n(\Mb)=\lambda_{\max}(\Mb)$ be its ordered eigenvalues.  $\mathrm{Ker}(\Mb)$ ($\mathrm{Ran}(\Mb)$) is the kernel (range) of $\Mb\in\mathbb{R}^{n\times n}$. $\sigma(\Mb)$ denotes the spectrum of  $\Mb\in\mathbb{R}^{n\times n}$. We use $\vb^\top$ to denote the transpose of a real vector $\vb$ and $\vb^H$ to denote the conjugate transpose of a complex vector $\vb$. For two matrices $\Mb_1$ and $\Mb_2$ having compatible dimensions, $\Mb_1\geq \Mb_2$ ($\Mb_1> \Mb_2$) means that $\Mb_1-\Mb_2$ is positive semi-definite (positive definite).  $\otimes$ is Kronecker product. We use $\Re(\cdot)$ to represent the real part of a complex number and $\mathcal{CLO}(\cdot)$ to denote the closure of a set.
$f\in L^p(0,T;\mathbb{R})$ means that $f:[0,T]\to\mathbb{R}$ is measurable \cite{Rudin} and 
$
\|f\|_{L^p(0,T)}=(\int_0^T |f|^p\mathrm{d}\mu)^{1/p}\leq C,
$ 
where  $C$ is a positive constant.
Moreover,  $f_k(t)\to f^*(t)$ as $k\to\infty$ if and only if 
$
(\int_{0}^{T}|f_k(t)-f^*(t)|^p\mathrm{d}t)^{1/p}\to 0$ as $ k\to\infty,
$ 
where $f_k,f^*\in L^p(0,T;\mathbb{R})$.

\medskip

\textit{Graph Basics}: The following concepts are adopted from \cite{Algebraic-graph-theory,Wu-synchronization}. The interactions between linear systems are characterized by an undirected dynamic graph $\mathcal{G}(t)=(\mathcal{V},\mathcal{E}(t),  \Wb(t))$ in which:
\begin{itemize} 

\item $\mathcal{V}=\left\{ 1,2,\ldots ,N\right\}$ is the set of nodes, each representing a single linear system;

\item $\mathcal{E}(t)\subset\mathcal{ V\times V}$ represents an edge set according to the following convention: $(i,j)$ belongs to $\mathcal{E}(t)$ if the information of node $i$ is available to node $j$ at time $t$.
	
\item $\Wb(t)=[ w_{ij}(t)] \in \mathbb{R}^{N\times N}$ is  adjacency matrix, where each $w_{ij}(t)=w_{ji}(t)$ is the weight of the
edge $(j,i)$ at time $t$.

\end{itemize} 
Moreover, $w_{ij}(t)=w_{ji}(t)>
0$ if $(j,i)$ is an edge of $\mathcal{G}(t)$ and $w_{ij}(t)=0$ otherwise.
Let $w_{ii}(t)=0$ for all $ t\geq 0$ and all $i\in \mathcal{V}.$ The Laplacian matrix $\Lb(t)$ of
$\mathcal{G}(t)$ is defined as
$\Lb(t)=\mathrm{diag}\{\Delta_{1}(t),\ldots,\Delta_{N}(t)\}-\Wb(t),$ where $\Delta_{i}(t)=\sum\nolimits_{j=1}^{N}w_{ij}(t)$ is the degree of node $i,$ $i=1,\ldots,N$.  
A path of length $r$ from $i_1\in\mathcal{V}$ to $i_q\in\mathcal{V}$ is a sequence of $r+1$ distinct vertices of the  form $(i_{1,}i_{2}),$ $(i_{2,}i_{3}),\ldots,
(i_{r},i_{r+1})\in\mathcal{E}(t).$ An undirected graph is connected if any two distinct nodes are connected to each other by at least one path. Note that even if $\mathcal{G}(t)$ is not connected at a particular instant of time, so that $\Lb(t)$ has more than one linearly independent nullvector, one can have consensus if $\mathcal{G}(t)$ is jointly connected (the term being defined later).

\section{Problem of Interest and Preliminary Observations \label{sec:problem formulation}}

\subsection{System Model and Problem of Interest}
Consider the following $N$ partial-state coupled linear systems
\begin{align}\label{linear-system-dynamics}
\dot{\xb}_i(t)=\Ab\xb_i(t)+ \Bb\Kb\sum_{j=1}^N w_{ij}(t)\left(\xb_j(t)-\xb_i(t)\right)
\end{align}
for $i=1,\ldots,N$,  where $\xb_i$ is the state of the $i$-th agent, $\Ab\in\mathbb{R}^{n\times n}$ and $\Bb\in\mathbb{R}^{n\times m}$ are, respectively, the system matrix and input matrix, being common to all $N$ systems.  We require $\Re(\lambda(\Ab))\geq 0$ for all $\lambda(\Ab)\in\sigma(\Ab)$\footnote{This can be relaxed such that there exists at least one $\lambda(\Ab)\in\sigma(\Ab)$ satisfying $\Re(\lambda(\Ab))\geq 0$. (Of course, if $\Ab$ is Hurwitz, the problem is trivial and of no interest).  The main results hold with controllability and observability being replaced by stabilizability and detectability, respectively. We make this requirement to keep the analysis concise.}. $\Kb\in\mathbb{R}^{m\times n}$ is the feedback matrix to be designed.

Define $\eb_i=\xb_i-\frac{1}{N}(\mathbf{1}_N^\top\otimes\Ib_n) \xb$ for $i=1,\ldots,N$, where $\xb=[\xb_1^\top,\ldots,\xb_N^\top]^\top$. If $\eb_i=\zeb$ for all $ i\in\mathcal{V}$, then (average) consensus is reached. Clearly, $\eb=(\Jb\otimes \Ib_n)\xb$ where $\eb=[\eb_1^\top,\ldots,\eb_N^\top]^\top$ and $\Jb=\Ib_{nN}-\frac{1}{N}\mathbf{1}_N\mathbf{1}^\top_N\otimes \Ib_n$. An important fact is $\eb^\top (\mathbf{1}_N\otimes \Ib_n)=\zeb.$ The evolution of $\eb$ can then be described by
\begin{align}\label{error-dynamics}
\dot{\eb}(t)=\left(\Ib_N\otimes \Ab-\hat{\Lb}(t)\otimes \Bb\Kb\right)\eb(t),
\end{align}
where $\hat{\Lb}(t)=\Lb(t)+\frac{1}{N}\mathbf{1}_N\mathbf{1}_N^{\top}$.

We first introduce the concept of global uniform exponential consensus for  system \eqref{linear-system-dynamics}.

\smallskip
\begin{definition}[Global  Uniform  Exponential Consensus]\label{exponential-consensus}The linear system \eqref{linear-system-dynamics} is said to achieve  global uniform exponential consensus (GUEC) if there exist $\gamma_1,\gamma_2>0$ such that along the trajectory of \eqref{linear-system-dynamics}, there holds $ \|\Phi(t,s)\|\leq \gamma_1e^{-\gamma_2 (t-s)}$ for all $t\geq s\geq 0$, where $\Phi(t,s)$ is the state transition matrix of system \eqref{error-dynamics}.
\end{definition}

We would like to point out that according to Definition \ref{exponential-consensus},   consensus will be achieved for all initial states.

\smallskip

\textit{Problem of Interest}. Design a suitable feedback matrix $\Kb$ and  find the weakest possible   conditions for GUEC of linear  system \eqref{linear-system-dynamics}.  

\subsection{Analysis Framework\label{subsec:analysis framework}}

Let $V(\eb)=\eb^{\top}(\Ib_N\otimes \Pb)\eb=\sum_{j=1}^N\eb_i^\top\Pb\eb_i$ be the Lyapunov function candidate for system \eqref{linear-system-dynamics},   where $\Pb>0$ is  to be determined. Note that    $V(\eb)$ is a sum of terms of the same functional form, with each summand reflecting one of the individual subsystems. While such a choice is undeniably restrictive, it allows the formulation of intuitively appealing conditions for stability of the complete system.  For $V$ (for brevity, we use $V$ instead of $V(\eb)$ without causing confusion), one has
\begin{align}\label{temp-4}
\lambda_{\min}(\Pb)\|\eb\|^2\leq V\leq \lambda_{\max}(\Pb)\|\eb\|^2.
\end{align}

Choosing $\Kb=\Bb^\top\Pb$  for system \eqref{linear-system-dynamics}, the evolution of $V$ along \eqref{error-dynamics} is governed by
\begin{equation}\label{tem-eqn-3}
\begin{split}
\dot{V}\big|_{\eqref{error-dynamics}}&=\eb^\top \left[\Ib_N\otimes(\Ab^\top\Pb+\Pb\Ab)-2\hat{\Lb}(t)\otimes \Pb\Bb\Bb^\top \Pb\right]\eb    \\ 
&\triangleq -\alpha(t) \eb^\top (\Ib_N\otimes \Pb)\eb=-\alpha(t)V.
\end{split}
\end{equation}
The term $\alpha(t)$ is given as follows:
\begin{align}\label{alpha-expression}
\alpha(t)=\frac{-\eb^\top \left[\Ib_N\otimes(\Ab^\top\Pb+\Pb\Ab)-2\hat{\Lb}(t)\otimes \Pb\Bb\Bb^\top \Pb\right]\eb}{\eb^\top (\Ib_N\otimes \Pb)\eb}
\end{align}
where $\eb(t)$ is assumed to be \textit{nonzero at any finite time} without loss of generality (To have $\eb(t)=\zeb$ at some finite time would mean, since $\eb(t)$ obeys a linear differential equation, that $\eb(0)=\zeb$, in which case consensus is achieved from the start.) The assumption on $\eb(t)$ ensures that $\alpha(t)$ is well defined. It follows from  \eqref{temp-4} and the concept of state transition matrix that GUEC (see Definition \ref{exponential-consensus}) is reached for system \eqref{linear-system-dynamics} if and only if  system \eqref{tem-eqn-3} is globally uniformly exponentially stable (GUES), i.e., there exist $\gamma_3,\gamma_4>0$ such that along the trajectory of \eqref{linear-system-dynamics}, there holds $\Phi_V(t,s)\leq \gamma_3e^{-\gamma_4 (t-s)}$ for all $t,s\geq 0$, where $\Phi_V(t,s)=\exp\{\int_s^t -\alpha(\tau)\mathrm{d}\tau\}$ is the state transition matrix of system \eqref{tem-eqn-3}.

The following  lemma presents a necessary and sufficient condition for GUES of system \eqref{tem-eqn-3}.
For clarity of presentation, we defer the proof of Lemma \ref{convergence-lemma} (and also proofs of the remaining lemmas and theorems) to Section \ref{sec:proofs}.

\begin{lemma}\label{convergence-lemma}
Consider system \eqref{tem-eqn-3}. Suppose that $\alpha(t)$ is measurable and bounded above by $\alpha^*>0$. Then, the existence of $a,T>0$ such that $ \int_t^{t+T}\alpha(s)\mathrm{d}s\geq a$ for all $t\geq 0$ is necessary and sufficient for GUES of system \eqref{tem-eqn-3}.
\end{lemma}

Note that if $\alpha(t)\ge 0$, then Lemma \ref{convergence-lemma} is a direct corollary of \cite[Theorem 1]{TAC1977}.
It can be easily observed from Lemma \ref{convergence-lemma} that the existence of a positive lower bound on the integral $\int_t^{t+T}\alpha(s)\mathrm{d}s,\;\forall t$ with some fixed $T>0$ is crucial for GUES of system \eqref{tem-eqn-3}.  As a result, we shall   characterize such a bound in the rest of this paper.   

\subsection{Assumptions \label{sec:problem formulation_assumptions} and Definitions}

\begin{figure*}[t]
	\hrule
	\begin{align}\label{integral-form-alpha}
	\int_{t}^{t+T}\alpha(s)\mathrm{d}s
	=-\int_{t}^{t+T}\frac{\eb^\top(t) \Phi^\top(s,t)\left[\Ib_N\otimes(\Ab^\top\Pb+\Pb\Ab)-2\hat{\Lb}(s)\otimes \Pb\Bb\Bb^\top \Pb\right]\Phi(s,t)\,\eb(t)}{\eb^\top(t) \Phi^\top(s,t)(\Ib_N\otimes \Pb)\Phi(s,t)\,\eb(t)}\mathrm{d}s
	\end{align}
	\hrule
\end{figure*}

We first present two weak assumptions. They are generalization of  various  interesting assumptions that are usually considered from different perspectives, but addressed by us in the one framework.

\begin{assumption}\label{boundedness}
$w_{ij}(t)$ are  measurable functions for all $i,j\in\mathcal{V}$. Moreover, there exists a $w^*>0$ such that	$0\leq w_{ij}(t)\leq w^*$ for all $t\geq 0$ and all $i,j\in\mathcal{V}$.   
\end{assumption}
\begin{remark}  
Assumption  \ref{boundedness} is quite mild and allows $w_{ij}(t)$ to be continuous or piecewise constant. Assumption \ref{boundedness} also guarantees that $\alpha(t)$ is measurable and bounded.
\end{remark}

For arbitrary but fixed $T>0$, and given any $r\geq0$, let $\hat{\Lb}_{r}(s):\mathbf{R}\to\mathbf{R}^{N\times N}$ be a matrix-valued function defined on $[0,T]$ such that $\hat{\Lb}_{r}(s)=\hat{\Lb}(s+r),s\in[0,T]$. Define $\Sigma$ by
	\begin{align*}
		\Sigma=\Big\{\hat{\Lb}_{r}(s),s\in[0,T]\big| r\geq 0\Big\}.
\end{align*}
\begin{assumption}\label{compactness-edge-weight}
With $\Lb(t)$ the Laplacian matrix of the graph $\mathcal{G}(t)$,     there exists $T>0$ such that $\Sigma$ is precompact\footnote{A precompact set is a set whose closure is compact \cite{Rudin}.}.
\end{assumption}

\begin{remark}\label{rmk-motivation}
Assumption \ref{compactness-edge-weight} characterizes the property of $\mathcal{G}(t)$ over intervals of a fixed length, which follows the idea of defining joint connectivity (see Definition \ref{joint-connectivity}). This assumption aims to unify and generalize existing conditions on how $\mathcal{G}(t)$ varies, e.g.,  piecewise continuous/constant condition \cite{QinTNNLS} (see Section \ref{sec: compact set of functions} for more discussions).  Based on this assumption, we are able to analyze GUEC of system \eqref{linear-system-dynamics} when $\mathcal{G}(t)$ is jointly connected and piecewise continuous (with  even infinite discontinuities on bounded intervals), which is left to be an open problem in existing literature. 
\end{remark}

\begin{remark}
	Roughly speaking, for $\Sigma$ to be precompact, $w_{ij}(t)$ cannot vary too fast for all $i,j\in\mathcal{V}$. An example of a function $w_{ij}(t)$ which should not be included is $\sin(\log|\sin(t)|)$. Intuitively, if the function contains sufficiently high frequency  it will be useless in the achievement of exponential convergence. The precompactness of $\Sigma$ is used to ensure the existence of the lower   bound (see  \eqref{tem-1} in the proof of Theorem \ref{unstable-linear-systems}) for the convergence to consensus of system \eqref{linear-system-dynamics}.
\end{remark}

\begin{remark}\label{generalization-rmk}  
	Assumption  \ref{compactness-edge-weight} generalizes those of existing literature imposed on network topology. Below in Section \ref{sec: compact set of functions} we identify certain classes of functions which assure the precompactness property. For example,   we show that if $\mathcal{G}(t)$ is piecewise constant and has a positive   dwell time, then $\Sigma$ is precompact; and if $\mathcal{G}(t)$ changes continuously such that $w_{ij}(t)$ are uniformly continuous on $[0,+\infty)$, then $\Sigma$ is also precompact. 
\end{remark}

\begin{remark}\label{rmk-1}
It is worthwhile to point out that if $\Lb(t)$ is a periodic function, i.e., $\exists T^0>0$ such that $\Lb(t)=\Lb(t+T^0)$ for all $t$, then $\Sigma$ can be defined as
	$\Sigma=\{\hat{\Lb}(s),s\in[0,T^0]\}.$ That is to say, $\Sigma$ consists of only one element which is a matrix-valued function defined on $[0,T^0]$. In this case,  $\Sigma$ is obviously precompact. Moreover, the results  in Section \ref{sec: necessary conditions} can be obtained via similar arguments by restricting the analysis on the time interval $[kT^0,(k+1)T^0]$ for any integer $k\geq 0$. 
\end{remark}

A further discussion of Assumption \ref{compactness-edge-weight} is given as follows. 
Write the integral of $\alpha(t)$ over $[t,t+T]$  in the form of  \eqref{integral-form-alpha}   with the aid of the state transition matrix $\Phi(t,0)$ of system \eqref{error-dynamics}. Inspired by Lemma \ref{convergence-lemma}, one needs to analyze the lower  bound of  \eqref{integral-form-alpha}.
To this end, we will investigate the  lower and upper bounds  of matrices $\Fb_i(t),\,i=1,\cdots,4$, which are defined as follows:
\begin{align}\label{F-function}
\Fb_i(t)= \int_{t}^{t+T}\Phi^\top(\tau,t) \Gamma_i\Phi(\tau,t)\mathrm{d}\tau
\end{align}
where $\Gamma_1=\Ib_N\otimes(\Ab^\top\Pb+\Pb\Ab)$, $\Gamma_2(t)=\hat{\Lb}(t)\otimes \Pb\Bb\Bb^\top \Pb$,  $\Gamma_3=\Ib_N\otimes \Pb$, and $\Gamma_4=\Ib_N\otimes \Pb\Bb\Bb^\top \Pb$.
For this aim, we collect a family of matrices, which is denoted by $\Sigma$ in Assumption \ref{compactness-edge-weight}.  The precompactness of $\Sigma$ is critical for assuring the existence of the desired lower bound, which will be shown in the forthcoming analysis.

Next, we introduce two definitions from graph theory, which are used in subsequent analysis. 
\begin{definition}[Union of Graph \cite{AndersonTAC2017}]
The union of the dynamically changing graph $\mathcal{G}(t)$ across $[t_0,t_1)$ is a graph with the same node set $\mathcal{V}$,  the adjacency matrix $\bar{\Wb}=[\bar{a}_{ij}]$ satisfying $\bar{a}_{ij}=\int_{t_0}^{t_1} w_{ij}(\tau)\mathrm{d}\tau$, and the edge set $\bar{\mathcal{E}}$ induced from $\bar{\Wb}$.
\end{definition}

\begin{definition}[Joint $(\delta,T)$-connectivity \cite{AndersonTAC2017}]\label{joint-connectivity}
The dynamic graph $\mathcal{G}(t)$ is said to be jointly $(\delta,T)$-connected if there exist positive real numbers $\delta$ and $T$ such that the edges 
$
(j,i):\;\;\int_t^{t+T}w_{ij}(\tau)\mathrm{d}\tau\geq \delta
$
form an undirected connected graph (also termed $\delta$-graph) over the node set $\mathcal{V}$ for all $t\geq 0$.
\end{definition}
          
\section{A Precompact Set of Laplacian Matrix-valued Functions\label{sec: compact set of functions}}

In this section,  we provide a few intuitive conditions to illustrate the  essence of Assumption \ref{compactness-edge-weight} and to show when Assumption \ref{compactness-edge-weight} is satisfied.  These conditions are derived via Simon’s famous general
result on the compact sets in the space $L^p[0,T]$ \cite{Simon} and are of independent interest.

In view of Remark \ref{rmk-1}, we only consider the case that $\Lb(t)$ is \textit{not} periodic in what follows. Since the Laplacian matrix $\Lb(t)$ has finite elements, we can construct  a set  of real-valued functions   from $w_{ij}(t)$ for any fixed $i,j \in\mathcal{V}$.    The set, denoted by $\Sigma^{ij}_1$, is formally defined as follows:
$
\Sigma^{ij}_1=\{ f_r(t),t\in[0,T]|   
 r\geq0\},
$
where $f_r(s)=w_{ij}(r+s),
s\in[0,T]$ for the given $r\geq0$.
Then, \textit{$\Sigma$ is precompact if and only if $\Sigma^{ij}_1$ are precompact} for all $i,j$. 
In light of this fact, we consider a real-valued   function $w(t)$ and define $\Sigma_1$ as $\Sigma_1^{ij}$ with respect to $w(t)$.  For simplicity, we will derive conditions in what follows for $\Sigma_1$ to be precompact.  \textit{It is worth pointing out that if $w_{ij}(t)$ satisfy the same properties that $w(t)$ has for all $i,j\in\mathcal{V}$, then  $\Sigma$ is precompact.}

Before presenting the first theorem, we introduce  some notations.   $\#(\mathcal{I})$ denotes the Lebesgue measure \cite{Rudin} of a set of real numbers $\mathcal{I}$. A real-valued function $h(t)$ is said to be continuous on $[t_1,t_2]$ if it is continuous on $(t_1,t_2)$, and is continuous from the right and left at $t_1$ and $t_2$, respectively. Let $\Omega$ be an arbitrary set of  discontinuous points of $w(t)$ on $[0,\infty)$ that has a zero Lebesgue measure. 

The following theorem says that if $w(t)$ does not change too fast in a certain sense, then $\Sigma_1$ is a precompact set.

\begin{theorem}\label{piecewise-constant-compactness}
Assume that   $w(t)$ is bounded. Let $g(t)$ denote the restriction of $w(t)$ to the set $[0,+\infty)\setminus\Omega$. Given any $t\in[0,\infty]\setminus\Omega$,  let $\mathcal{I}_t$ be the largest open interval containing $t$ such that $g$ is continuous on $\mathcal{I}_t$. Suppose there exist  positive real numbers $c,\,\hat{c}>0$ such that given any interval $[s,p]$, 
	\begin{enumerate} 
		\item[(i)] if $g$ is continuous on the interval $[s,p]$, then $|g(s)-g(p)|\leq c|s-p|$; 
		\item[(ii)] otherwise, $|g(s)-g(p)|\leq \hat{c}[\#(\mathcal{I}_s)+|s-p|]$  when  $\mathcal{I}_s$ is not empty; or $|g(s)-g(p)|\leq \hat{c}(s-p)$  when  $\mathcal{I}_s$ is empty.
	\end{enumerate} 
	then $\Sigma_1$ is a precompact set.
\end{theorem}

\begin{remark}
Condition (i) requires that $g(t)$ has at most a linear growth when it is continuous. This condition precludes any isolated essential discontinuity \cite{Rudin}. For instance, the function $\sin(1/(t-1))$ does not satisfy condition (i) when $t$ approaches $1$ from the left.  Condition (ii) restricts the change of $g(t)$ at a jump discontinuous point to the length of a continuous interval associated with this point. The set $\Omega$ is used to remove points (or equivalently redefine the values of these points) with the result that $g(t)$ behaves well on $[0,+\infty)\setminus\Omega$.
\end{remark}
  
\begin{remark}
Theoretically, Theorem \ref{piecewise-constant-compactness} includes  functions $w(t)$ that have an infinite number of discontinuities in a bounded interval. Let $w(t)$ be defined from a Lipschitz continuous function $h(t)$ such that $w(t)=0$ whenever $t$ is a rational number and $w(t)=h(t)$ otherwise.  By choosing $\Omega$ to be rational numbers on $[0,+\infty)$, $g(t)$ is Lipschitz continuous on $[0,+\infty)\setminus\Omega$ and $\Sigma_1$ is precompact according to  Theorem \ref{piecewise-constant-compactness}.  
\end{remark}

Next we present two interesting corollaries. It is shown that the conditions frequently used in existing literature are special cases of Assumption \ref{compactness-edge-weight}.

\begin{corollary}\label{positive-dewell-time}
 Assume  that    there exist nonempty and contiguous intervals  $[s_{0},s_{1}),[s_{1},s_{2}),\ldots,[s_{k},s_{k+1}),\cdots$ such that $\cup_js_{j}=[0,\infty)$ and $w(t)$ is continuous on each interval $(s_k,s_{k+1})$. Suppose that the following conditions hold:
 \begin{itemize}
 	\item[(i)] $|w(t_1)-w(t_2)|\leq c|t_1-t_2|$ if $t_1,t_2\in[s_j,s_{j+1})$ for some $j\geq 0$,
 	\item[(ii)] $|w(s_j^+)-w(s_j^{-})|\leq \hat{c}|s_{j-1}-s_j|$ for $j\geq 1$, where $w(s_j^+)$ and $w(s_j^{-})$ denote the limits from the right and left at $s_j$, respectively,
 \end{itemize}
  then $\Sigma_1$ is  precompact.
\end{corollary}

Note that we allow $\inf_{k\in \mathbb{Z}_+} (s_k-s_{k-1})$, which is termed as dwell time, to be zero in Corollary \ref{positive-dewell-time}. In other words, $w(t)$ can have an infinite number of discontinuities in a bounded interval.  Corollary \ref{positive-dewell-time} also gives a useful principle for designing $\mathcal{G}(t)$. If the network topology switches fast inevitably, the design of $\mathcal{G}(t)$ such that Assumption  \ref{compactness-edge-weight} holds, together with other mild conditions, ensures exponential consensus (see Section \ref{sec: sufficient conditions}) and provides more tolerance to disturbances.  
 
If $w(t)$ is  constant on any $[s_j,s_{j+1})$ and $\inf_{k\in \mathbb{Z}_+} (s_k-s_{k-1})>0$, then the condition can be further simplified for $\Sigma_1$ to be precompact.

\begin{corollary}\label{compactness-no-dewell-time}
 Assume that    $w(t)$ is bounded and piecewise constant. Let $[s_{0},s_{1}),[s_{1},s_{2}),\ldots,[s_{k},s_{k+1}),\cdots$ be a sequence of nonempty and contiguous intervals such that $\cup_{j=0}^{\infty}[s_{j},s_{j+1})=[0,\infty)$ and $w(t)$ is constant on each $[s_{j},s_{j+1})$. If $\inf_{k\in \mathbb{Z}_+} (s_k-s_{k-1})>0$, then $\Sigma_1$ is a precompact set.
\end{corollary}

If $w(t)$ is a  continuous function, then $\Sigma_1$ is also precompact when $w(t)$ is uniformly continuous, even if the derivative of $w(t)$ may grow unboundedly.

\begin{lemma}\label{uniform-compactness}
Suppose that  $w(t)$ is bounded and uniformly continuous\footnote{A real-valued function $g$ is uniformly continuous if for every real number $\epsilon>0$, there exists a $\delta(\epsilon)>0$ such that $|g(x_1)-g(x_2) |\leq \epsilon$ whenever $|x_1-x_2 |\leq \delta(\epsilon)$ \cite{Rudin}.} on $[0,+\infty)$. Then, the set $\Sigma_1$ is precompact.
\end{lemma}

\begin{remark}
We note that if $D^{+}w(t)$ is bounded everywhere, then $w(t)$ is uniformly continuous (where $D^{+}$ denotes the Dini derivative 
\cite{Hagood}).	
In particular, if $w(t)$ is Lipschitz continuous, then $D^{+}w(t)$ is bounded everywhere.
\end{remark}

\section{Necessary and Sufficient Conditions for Exponential Consensus\label{sec: necessary conditions}}

\subsection{Necessary Conditions for Exponential Consensus\label{necessity}}
In this section, we analyze the necessity of joint $(\delta,T)$-connectivity of $\mathcal{G}(t)$  and   controllability of $(\Ab,\Bb)$ for GUEC of system \eqref{linear-system-dynamics} under Assumptions \ref{boundedness} and \ref{compactness-edge-weight}. The necessity seems intuitive, however its proof is technically challenging.
We first recall a test of controllability \cite{ChenLinearSystem}.

\begin{lemma}[cf.\cite{ChenLinearSystem}]\label{negative-definite-A}
$(\Ab,\Bb)$ is controllable if and only if the controllability matrix $[\Bb,\Ab\Bb,\ldots,\Ab^{n-1}\Bb]$ is of full-row rank. 
\end{lemma}

\begin{remark}\label{rmk-2}
A matrix pair $(\Ab,\Bb)$ being not controllable is equivalent to the existence of $\vb\neq\zeb$ such that $\vb^H \Ab^i\Bb=\zeb$ for all $i=0,\ldots,n-1$ and this is in turn equivalent to the existence of  $\vb\neq\zeb$ such that  $\vb^H \Ab=\lambda \vb^H$   and $\vb^{H}\Bb=\zeb$ \cite{ChenLinearSystem}.
\end{remark}

The following theorem, which is not altogether surprising, says that  the controllability of  $(\Ab,\Bb)$ is necessary for GUEC of system \eqref{linear-system-dynamics} with any feedback matrix $\Kb$ and any $\mathcal{G}(t)$.

\begin{theorem}\label{necessity-controllability}
	Consider the linear interconnected system \eqref{linear-system-dynamics} communicating over $\mathcal{G}(t)$.   If consensus is achieved  globally, uniformly, and exponentially fast, then $(\Ab,\Bb)$ is controllable.
\end{theorem}

\begin{remark}
The necessity of the controllability of $(\Ab,\Bb)$ was investigated for consensus of multi-agent systems in \cite{Ma}. However the analysis relies on the invariance of controllability property under any equivalence transformation and only applies  to a fixed and connected communication graph. In contrast,  we relax the connectivity condition and show that  if $(\Ab,\Bb)$ is not controllable, then there exists a non-trivial subspace in which a vector cannot be controlled to leave this space, thus global consensus is impossible  by Remark \ref{rmk-2}.
\end{remark}

Lemma \ref{convergence-lemma} implies that with $V(\eb)$, GUEC for system \eqref{linear-system-dynamics} can be achieved if and only if  the integral of $\alpha(t)$ over any time interval with a fixed length $T$ has positive lower and upper bounds. By Eq. \eqref{integral-form-alpha}, Assumption \ref{compactness-edge-weight} and Lemma \ref{state-transition-matrix-depend-on-laplacian-matrix} in Section \ref{sec:proofs}, and noting that $\eb(t)$ can be any vector in $\mathbb{R}^{nN}$ since $\eb(0)$ is arbitrary, the above condition is equivalent to requiring that there exist positive real numbers $\alpha_1,\,\alpha_2$, and $T$ such that 
\begin{align}\label{temp-2}
\alpha_1\Ib\leq-\Fb_1(t)+2\Fb_2(t)\leq \alpha_2\Ib
\end{align}
holds for any $t\geq 0$. This condition \eqref{temp-2} does not depend on system state.

Next, we show  the necessity of joint $(\delta,T)$-connectivity of $\mathcal{G}(t)$ for \eqref{temp-2} to be satisfied with Assumptions \ref{boundedness}--\ref{compactness-edge-weight} and the choice of feedback matrix $\Kb=\Bb^\top\Pb$ for some positive  definite matrix $\Pb$. To show the necessity of joint $(\delta,T)$-connectivity of $\mathcal{G}(t)$,   we should relate joint connectivity to  exponential convergence of the states to consensus. To this end, we develop an analysis framework in Section II, which leads us to characterize the lower bound of the following integral:
\begin{align*}
	\int_t^{t+T}\Phi^\top(\tau,t)\left(\hat{\Lb}(\tau)\otimes \Pb\Bb\Bb^\top\Pb\right)\Phi(\tau,t)\mathrm{d}\tau.
\end{align*}
The above integral involves the term $\Phi(\tau,t)$. Thus, its lower bound is jointly determined by $\Ab$, $\Bb$, $\Pb$, and $\Lb(t)$  and is difficult to analyze.

\begin{theorem}\label{necessity-connectivity-neutralstability}
Consider the linear interconnected system \eqref{linear-system-dynamics} communicating over $\mathcal{G}(t)$. Suppose that Assumptions \ref{boundedness} and \ref{compactness-edge-weight} hold  and $\Kb=\Bb^\top\Pb$ for some $\Pb>\zeb$. 
If consensus is achieved  globally, uniformly, and exponentially fast, then $\mathcal{G}(t)$  is jointly $(\delta,T)$ connected.
\end{theorem}

The  joint $(\delta,T)$-connectivity condition is quite mild. It is weaker than the uniform joint connectivity with the existence of a positive dwell time  \cite{SuTAC2012,Wang}.
Moreover, if such a connectivity condition does not hold, then GUEC cannot be ensured in some cases (see Example \ref{exp-delta-T-connectivity} in Section \ref{sec:examples}). 
Finally,   the necessity of joint  connectivity of $\mathcal{G}(t)$ is proved based on the design  of   $\Kb=\Bb^\top\Pb$ and the quadratic Lyapunov function candidate $V(\eb)$.


\subsection{Sufficient Conditions for Exponential Consensus\label{sec: sufficient conditions}}

Under Assumptions \ref{boundedness}--\ref{compactness-edge-weight}, we show a sufficient condition   for system \eqref{linear-system-dynamics} to realize GUEC. In general, simply putting together the two necessary conditions in  Section \ref{necessity} cannot guarantee GUEC of system \eqref{linear-system-dynamics}. We need to further characterize  how system parameters and joint $(\delta,T)$-connectivity condition work together for reaching GUEC, which motivates the synchronization index mentioned in the Abstract and Introduction sections. 

\begin{theorem}\label{unstable-linear-systems}
Consider the linear interconnected system \eqref{linear-system-dynamics} communicating over $\mathcal{G}(t)$. Suppose that Assumptions \ref{boundedness} and \ref{compactness-edge-weight} hold, $\sigma(\Ab)$ belongs to the closed right-half plane. Design $\Kb=\Bb^\top\Pb$ with $\Pb$ satisfying 
\begin{align}\label{theorem-are}
\Ab^\top \Pb+\Pb\Ab-\kappa_1 \Pb\Bb\Bb^\top \Pb+\Qb=\zeb
\end{align}
for some $\kappa_1>0$ and $\Qb\geq \zeb$ where $(\Ab,\Qb^{1/2})$ is observable. Assume that  $\Fb_2(t)\geq \frac{\kappa_2}{2}\,\Fb_4(t)$ for some $\kappa_2>0$ and for all $t$.
If $(\Ab,\Bb)$ is controllable, $\mathcal{G}(t)$ is jointly $(\delta,T)$-connected, and $\frac{\kappa_2}{\kappa_1}\geq 1$, then consensus is achieved globally, uniformly, and exponentially fast.
\end{theorem}
\begin{remark}\label{rmk-3}
Since   $(\Ab,\Bb)$ is controllable and $(\Ab,\Qb^{1/2})$ is observable, for any $\kappa_1>0$, there exists a positive definite $\Pb$ such that \eqref{theorem-are} holds. With the matrix $\Pb$ given, the existence of $\kappa_2$ such that $\Fb_2(t) \geq \frac{\kappa_2}{2}\Fb_4(t)$ for all $t \geq 0$ is guaranteed as a result of Assumptions 1 and 2 (refer to the detailed proof of Theorem \ref{unstable-linear-systems}). Although $\kappa_1$ and $\kappa_2$ are ensured to exist,  we cannot guarantee   $\kappa_2/\kappa_1\geq 1$ because $\kappa_2$ depends on $\kappa_1$. The quantity $\kappa_2/\kappa_1$ is crucial for achieving consensus and it reflects how system parameters  and
 joint $(\delta,T)$-connectivity condition
work together to guarantee GUEC.
\end{remark}

\begin{remark}
An explicit convergence rate that can be directly computed is not available in Theorem \ref{unstable-linear-systems}. We only prove that a lower bound for the convergence rate exists in the proof of Theorem \ref{unstable-linear-systems}.
\end{remark}

\begin{remark}
($i$) If $\Ab$ is neutrally stable, then  $\Pb>\zeb$ can be chosen such that $\Ab^\top \Pb+\Pb\Ab=\zeb$ \cite{SuTAC2012}. The derivative of $V$ along the state trajectory of system \eqref{error-dynamics} gives
$\dot{V}=-\alpha(t)V$, where $\alpha(t)\geq 0$. In this case, $\kappa_1$ can be an arbitrary positive value since we do not have \eqref{theorem-are}, which easily ensures $\kappa_2/\kappa_1\geq 1$.
($ii$) In addition, if $\mathcal{G}(t)$ remains connected for all $t\geq 0$ and moreover $\hat{\Lb}(t)\geq \beta \Ib$ for some $\beta>0$, then $\kappa_2$ can be chosen to be $2\beta$ according to its definition. As such, any $\kappa_1$ that is less than  $2\beta$ yields $\kappa_2/\kappa_1\geq 1$. The above discussion shows that for some special cases, the controllability of $(\Ab,\Bb)$ and joint $(\delta,T)$-connectivity of $\mathcal{G}(t)$  are necessary and sufficient for exponential consensus.  For a general case,   $\kappa_2/\kappa_1>1$ is additionally required to secure exponential consensus, whose necessity, unfortunately, is not obtained.
\end{remark}

The quantity $\kappa_2/\kappa_1$  can be viewed as a synchronization index of system \eqref{linear-system-dynamics} over a  dynamic network topology (see Remark \ref{rmk-3}).  
Theorem \ref{unstable-linear-systems}  says that $\kappa_2/\kappa_1\geq 1$ guarantees GUEC of system \eqref{linear-system-dynamics} provided that $(\Ab,\Bb)$ is controllable and $\mathcal{G}(t)$ is jointly $(\delta,T)$-connected under Assumptions \ref{boundedness}--\ref{compactness-edge-weight}.

It is quite a challenge to analytically specify conditions to ensure $\kappa_2/\kappa_1\geq 1$ in a general case.  Nevertheless, one can try to find   $\Pb$ to guarantee $\kappa_2/\kappa_1\geq 1$ numerically. Specifically, first take $\gamma_k=1/k$ where $k\in\mathbb{Z}_{+}$. Then, solve the algebraic Riccati equation
\begin{align} \label{alg-are}
\Ab^\top \Pb_{\gamma_k}+\Pb_{\gamma_k}\Ab-\gamma_k \Pb_{\gamma_k}\Bb\Bb^\top \Pb_{\gamma_k}+\Ib_n=\zeb
\end{align}
to obtain $\Pb_{\gamma_k}$. Finally, calculate $\kappa_2(\gamma_k)$ with $\Pb_{\gamma_k}$ such that $\Fb_2(t)\geq (\kappa_2(\gamma_k)/2)\,\Fb_4(t)$ for all $t$. If there exists a $\kappa_2$ such that $\kappa_2(\gamma_k)\geq \kappa_2$ for all $k$, then any $\kappa_1$ satisfying $\kappa_2\geq \kappa_1$  yields that $\kappa_2/\kappa_1\geq 1$. We summarize this procedure in Algorithm \ref{alg-1} and illustrate it in a numerical example (Example \ref{exp-2} in Section \ref{sec:examples}).

\begin{algorithm}
	\caption{-- Finding  $\kappa_2$}
	\label{alg-1}
	\begin{algorithmic}
		\STATE {Initiate matrices $\Ab$ and $\Bb$, and set $k=1$} 
		\REPEAT 
		\STATE Solve \eqref{alg-are} to obtain $\Pb_{k}$ with $\gamma_k=\frac{1}{k}$;
		\STATE Calculate $\kappa_2(\gamma_k)$ using $\Pb_{k}$;
		\STATE $k=k+1$;
		\UNTIL{$\kappa_2(\gamma_k)$ converges or $k$ is sufficiently large}
	\end{algorithmic}
\end{algorithm}

Algorithm 1 provides a strategy to search for a pair $(\kappa_1,\kappa_2)$ such that the synchronization index, $\kappa_2/\kappa_1$, is not less than $1$. The algorithm is essentially a searching procedure along a certain direction.  As a result, the iterations carried out by the algorithm can be manually set according to prior knowledge or to achieve a better trade-off between cost of computation and size of search area.

If a converged $\kappa_2(\gamma_k)$ is not obtained and the algorithm stops due to large iteration steps,  then one can numerically calculate the lower bound of the obtained sequence $\{\kappa_2(\gamma_k)/\gamma_k\}$. If the lower bound is obtained with $\kappa_2(\gamma_j)/\gamma_j\geq 1$ for some $j$, the synchronization index is guaranteed to be not less than 1 with $\kappa_1=1/j$.

\section{Proofs of Main Results\label{sec:proofs}}
In this section, we provide complete proofs of the main results. For brevity's sake, in the following proof we  drop the subscript $r$ and the time interval $[0,T]$ used for the definition of $\Sigma$ in Assumption \ref{compactness-edge-weight} without causing any confusion hereafter and use $\hat{\Lb}^b(t)$ to denote an arbitrary element of $\Sigma$.

{\subsection{\textsc{Proof of Lemma \ref{convergence-lemma}}}

(Sufficiency.)	Consider $\dot{V}=-\alpha(t)V$, whose state transition matrix is $\Phi_V(t,s)=\exp\{\int_s^t-\alpha(t)(\tau)\mathrm{d}\tau\}$.
Since $ \int_t^{t+T}\alpha(\tau)\mathrm{d}\tau\geq a$, it follows that $\Phi_V(s+kT,s)\leq  \exp\{-ak\}$ where $k=\lfloor (t-s)/T\rfloor$. By the definition of $k$, $t\in[s+kT,s+kT+T]$. In view of the fact that $\alpha(t)$ is bounded above by $\alpha^*$, $\int_{s+kT}^{t}-\alpha(\tau)\mathrm{d}\tau$ has an upper bound $0$ and a lower bound $-\alpha^*T$. Hence, 
\begin{align*}
\Phi_V(t,s)&=\Phi_V(t,s+kT)\Phi_V(s+kT,s)\\
&\leq e^{\int_{s+kT}^{t}-\alpha(\tau)\mathrm{d}\tau}e^{-ak} 
\leq e^{-ak}.
\end{align*}
Noting that $ k\geq (t-s)/T-1$, it is further obtained that
$
\Phi_V(t,s)\leq e^{-a(t-s)/T+a }.
$
Let $\gamma_3=\exp\{a \}>0$ and $\gamma_4=a/T>0$. Consequently, 
$$
\Phi_V(t,s)\leq \gamma_3\exp\{-\gamma_4 (t-s) \}. 
$$  

(Necessity.)	By the  global  uniform  exponential convergence of $V$, one has 
$$
\Phi_V(t,s)=e^{-\int_{s}^{t}\alpha(\tau)\mathrm{d}\tau }\leq \gamma_3e^{-\gamma_4 (t-s) }.
$$
 Let	 $t=s+T$, where $T>0$ is independent of $s$.
This gives
\begin{align*}
\int_{s}^{s+T}\alpha(\tau)\mathrm{d}\tau\geq -\ln \gamma_3+\gamma_4 T,
\end{align*}
as desired. Here, $T>\ln \gamma_3/\gamma_4$ when $\ln\gamma_3$ is positive.
This finishes the proof.

\subsection{\textsc{Proofs of Theorem  \ref{positive-dewell-time}, Lemma \ref{uniform-compactness}, and Corollaries \ref{positive-dewell-time}-\ref{compactness-no-dewell-time}}}

To complete the proofs, we need the following lemma which provides a criterion to verify whether a set of real-valued functions is precompact. 

Let $T>0$ and $(S_h w)(t)=w(t+h),\,t\geq 0,$ be the shift operator. Note that if $w(t)$ is defined on $[0,T]$ for some $T>0$, then $(S_h w)(t)$ is defined on $[0,T-h]$ with $T\geq h\geq 0$.

\begin{lemma}[cf.  {\cite[Theorem 1]{Simon}}]\label{compact-family-functions}
Let $F\subset L^p(0,T;\mathbb{B})$, where $\mathbb{B}$ is Banach space. $F$ is relatively compact\footnote{A relatively compact set, also called precompact set,   is a set whose closure is compact.} in $L^p(0,T;\mathbb{B})$ for $1\leq p<\infty$, or in $C(0,T;\mathbb{B})$ for $p=\infty$ if and only if 
\begin{numcases}{}
\left\{\int_{t_1}^{t_2}f(t)\mathrm{d}t: \; f\in F \right\} \text{ is relatively compact in } \mathbb{B},  \notag \\
\qquad\qquad\qquad\qquad\qquad\qquad \forall\, 0<t_1<t_2<T \label{temp-6}\\
\|f(t)-S_h f(t)\|_{L^p(0,T-h;\mathbb{B})}\to 0, \text{ as } h\to 0,\;  \notag \\ 
\qquad\qquad\qquad \text{uniformly for } f\in F. \label{bounded-variation}
\end{numcases}
\end{lemma}

\medskip

For the purpose of this paper, we simply take $\mathbb{B}=\mathbb{R}$. Write $L^p(0,T;\mathbb{R})$ as $L^p(0,T)$ for brevity.
Without causing confusion, we also use $L^p$ instead of $L^p(0,T)$. Moreover, let $p=1$, i.e., we consider the space $L^1(0,T)$.

\medskip

\textit{\textbf{Proof of Theorem \ref{positive-dewell-time} using Lemma \ref{compact-family-functions}}.}
By Assumption \ref{boundedness}, there exists a $C>0$ such that $-C\leq \int_{t_1}^{t_2}f(t)\mathrm{d}t\leq C$ for all $0<t_1<t_2<T$ and uniformly for $f\in\Sigma_1$.   As a result, the first condition, i.e., condition \eqref{temp-6}, in Lemma \ref{compact-family-functions} is satisfied. Then, to complete the proof, it suffices to show that for any $\epsilon>0$, there exists an $\eta>0$ such that 
$
\left\|f(t)-S_h f(t)\right\|_{L^1(0,T-h)}\leq \epsilon,\;\forall  h\leq \eta,
$
uniformly for $f\in\Sigma_1$. To this end, we first calculate $f(t)-S_hf(t)$ on $[0,T-h]$ in what follows for an arbitrary $f(t)\in\Sigma_1$ and $T>h>0$.

Let  $\Omega_f\subset\Omega$ be the set of points of discontinuity of $f$ on $[0,T-h]$. Now, for brevity, we construct a function $g_f$  that is defined on $[0,T-h]\setminus\Omega_f$ such that $g_f(s)=f(s)$ for any $s\in[0,T-h]\setminus\Omega_f$. Clearly, $g_f$ is a restriction of $f$ on $[0,T-h]\setminus\Omega_f$.
	
Let $t\in[0,T-h]$ be arbitrary and $g_f$ be defined at $t$ and $t+h$.  If  $g_f$ is continuous on $[t,t+h]\setminus\Omega_f$, then by the conditions in the theorem statement  and the construction of $\Sigma_1$, one has $g_f(t)-g_f(t+h)\leq ch$. As a result, there holds that 
\begin{align*} 
\left|f(t)-S_{h}f(t)\right|=\left|f(t)-f(t+h)\right|\leq c h.
\end{align*}

Now consider the scenario that  $g_f$ is not continuous on $[t,t+h]\setminus\Omega_f$.  Let $\mathcal{I}_t$ be the largest open interval containing $t$ such that $g_f$ is continuous on $\mathcal{I}_t$.

\begin{itemize}
\item \textsc{Case I: $\mathcal{I}_t$ is empty.} This implies that the function $g_f$ is not continuous on any open interval of the form $(s_1,s_2)$ with $s_1<t< s_2$. By the conditions in the theorem statement, one has $f(t)-f(t+h)=g_f(t)-g_f(t+h)\leq \hat{c} h.$

\item \textsc{Case II: $\mathcal{I}_t$ is not  empty}. In this case, 
 $f(t)-f(t+h)=g_f(t)-g_f(t+h)\leq \hat{c}\,\#(\mathcal{I}_t)+\hat{c}h,$
where $t$ satisfies that 
$
t-\min_s\{s\in\mathcal{I}_t\}\leq h.
$
\end{itemize}

Note that in the second case, if $t\in\mathcal{I}_t$ and 
$
\max_s\{s\in\mathcal{I}_t\}-t> h,
$
there holds 
$$
f(t)-f(t+h)=g_f(t)-g_f(t+h)\leq ch.
$$
This indicates that although a term independent of $h$, i.e., $\hat{c}\,\#(\mathcal{I}_t)$, exists,  it appears for the function $g_f$ on a sub-interval of $\mathcal{I}_t$ having length of at most $h$. 

 Define the union of all such open intervals for each $t\in[0,T-h]\setminus \Omega_f$  by $\mathcal{I}$. Then, by the above calculation, one arrives at the following inequalities
\begin{align*}
&\|f(t)-S_h f(t) \|_{L^1(0,T-h)}\\
& \leq \int_{\Omega_f}|f(t)-f(t+h)|\mathrm{d}t+\int_{[0,T-h]\setminus \Omega_f} \max\{\hat{c},c\}h\,\mathrm{d}t\\
&~~~~+\sum_{\mathcal{I}_s\in\mathcal{I}}\int_{ \mathcal{I}_s^{'} \subset\mathcal{I}_s} \hat{c}\,\#(\mathcal{I}_s)\mathrm{d}t\\
& \leq  \max\{\hat{c},c\}h(T-h)+\hat{c}h(T-h)\leq \max\{\hat{c},c\}Th,
\end{align*}
where we have used the fact that $\#(\Omega_f)=0$, $f(t)$ is bounded by $w^*$, $\#(\mathcal{I}_s^{'})\leq h$, and $\sum_{\mathcal{I}_s\in\mathcal{I}}\#(\mathcal{I}_s)\le T-h$. 
Therefore,  given any $\epsilon>0$, there exists an $\eta=\frac{\epsilon}{\max\{\hat{c},c\}T}$ such that 
$$
\left\|f(t)-S_h f(t) \right\|_{L^1(0,T-h)}\leq \epsilon,\;\forall f\in \Sigma_1,\;\forall  h\leq \eta.
$$
Here, the quantities $c,\,\hat{c}$, and $T$ are independent of $f$ chosen from $\Sigma$.  This finishes the proof of Theorem \ref{positive-dewell-time}.

\textit{\textbf{Proofs of Corollaries \ref{positive-dewell-time}-\ref{compactness-no-dewell-time}.}} It suffices to show that the conditions in Theorem \ref{positive-dewell-time} hold.

(1) Choose $\Omega=\emptyset$. Then, $g=w$ on $[0,+\infty)$.  It is clear that $|w(s)-w(p)|\leq c|s-p|$ with  $s,p\in [s_j,s_{j+1})$ for any $j\geq 0$. If there exist discontinuous points in $[s,p]$, then write $[s,p]=[s,s_{k+1})\cup\cdots\cup[s_\ell,p]$ where $[s,s_{k+1})\subset[s_k,s_{k+1})$,  $[s_\ell,p]\subset [s_\ell,s_{\ell+1})$ and $\ell> k+1$. Consequently,
\begin{align*}
	&|w(s)-w(p)|
	\leq \left|w\left(s^-_{k+1}\right)-w\left(s\right)\right|\\
	&+\left|w\left(p\right)-w\left(s_{\ell}\right)\right|
	+ \sum_{j=2}^{\ell-k}\left|w\left(s_{k+j}\right)-w\left(s_{k+j-1}\right)\right|\\
&	\leq \left|w\left(s^-_{k+1}\right)-w(s)\right|+\left|w(p)-w\left(s_{\ell}\right)\right|+\left|w(s^-_{k+1})-w\left(s_{k+1}\right)\right|\\
	&+ \sum_{j=2}^{\ell-k}\left(\left|w\left(s^-_{k+j}\right)-w\left(s_{k+j-1}\right)\right|+\left|w\left(s_{k+j}\right)-w\left(s^-_{k+j}\right)\right|\right)\\
	&\leq 2 \max\{c,\hat{c} \}\left[\#(\mathcal{I}_s)+|s-p|\right],
\end{align*}
which proves Corollary \ref{positive-dewell-time}.

(2) Since $\inf_{k\in \mathbb{Z}_+} (s_k-s_{k-1})>0$ and $w(t)$ is bounded, the second condition in Corollary  \ref{positive-dewell-time} holds.  Then, Corollary \ref{compactness-no-dewell-time} follows from Corollary  \ref{positive-dewell-time}.

\textit{\textbf{Proof of Lemma \ref{uniform-compactness} using Lemma \ref{compact-family-functions}. }}
For uniformly continuous and bounded $w(t)$, given any $\epsilon>0$, there exists $\eta>0$ such that  $|w(s)-w(p)|\leq \epsilon$ whenever $|s-p|\leq \eta$. Then, choosing $p=\infty$ in Lemma \ref{compact-family-functions}, condition \eqref{bounded-variation} holds uniformly for $f\in\Sigma_1$, which proves that $\Sigma_1$ is precompact.  

\subsection{\textsc{Proof of Theorem \ref{necessity-controllability}}}

Suppose that the matrix pair $(\Ab,\Bb)$ is not controllable. Then, there exists an eigenvector $\vb\in\mathbb{C}^n$ associated with the uncontrollable eigenvalue  $\lambda_u$   of $\Ab$, i.e., $\vb^H \Ab=\lambda_u \vb^H$ and $\vb^H \Bb=\zeb$ \cite{ChenLinearSystem}. For simplicity, let us assume that $\lambda_u$ is real and $\vb$ is also a real vector.

Let $\mathcal{C}$ be the controllability subspace determined by $(\Ab,\Bb)$, viz. $\mathcal{C}=\mathrm{Ran}([\Bb,\Ab\Bb,\cdots,\Ab^{n-1}\Bb])=\mathrm{Ran}(\exp(\Ab t)\Bb),\;t\geq 0$. Then, if  $\xb_i(0)\in \mathcal{C}$, $\xb_i(t)\in\mathcal{C}$ for all $t.$  This gives that if $\xb_i(0)\in\mathcal{C}$ for some $i\in\mathcal{V}$, then $\vb^\top \xb_i(t)=\vb^\top \xb_i(0)=0,\,\forall t$.

The evolution of $\vb^\top\xb_k(t)$ for any $k$ is expressible as follows:
\begin{align*}
\vb^\top\dot{\xb}_k=\vb^\top\Ab\xb_k+\vb^\top\Bb\Kb\sum_{j=1}^N w_{kj}(t)\left(\xb_j-\xb_k\right)
= \lambda_u\vb^\top\xb_k,
\end{align*} 
where we have used the fact that $\vb^\top\Bb=\zeb$ and $\vb^\top\Ab=\lambda_u\vb^\top$. Therefore,   $\vb^\top\xb_k(t)=e^{\lambda_u\cdot t}\vb^\top\xb_k(0).$

Let $\xb_i(0)\in \mathcal{C}$ for some $i$, then $\vb^\top\xb_i(t)=0$ for all $t\geq 0$. If consensus can be achieved for system \eqref{linear-system-dynamics} globally, then  $\vb^\top\xb_j(t)\to \zeb$ as $t\to\infty$ for all $j\neq i$ and any initial value $\xb_j(0)$. However, if $\vb^\top\xb_k(0)\neq 0$ for some $k\neq i$, then  $\vb^\top\xb_k(t)=e^{\lambda_u\cdot t}\vb^\top\xb_k(0)$, which clearly does not converge to zero. Hence, consensus cannot be achieved -- a contradiction This finishes the proof.  

\subsection{\textsc{Proof of Theorem \ref{necessity-connectivity-neutralstability}}}

The proof of Theorem \ref{necessity-connectivity-neutralstability} uses the following lemmas to characterize the eigenvalues of a Laplacian matrix.
\begin{lemma}[cf. \cite{Wu-synchronization}]\label{eigenvalue-bound}
Consider an undirected graph $\mathcal{G}$, whose Laplacian matrix is $\Lb$. Let $\mathcal{S}_1$ and $\mathcal{S}_2$ be two nontrivial disjoint subsets of $\mathcal{V}$, i.e., $|\mathcal{S}_1|>0$, $|\mathcal{S}_2|>0$, and $\mathcal{S}_1\cap\mathcal{S}_2=\emptyset$, $\mathcal{S}_1\cup\mathcal{S}_2=\mathcal{V}$.  Then, one has
$\lambda_2(\Lb)\leq e(\mathcal{S}_1,\mathcal{S}_2)/|\mathcal{S}_1|+e(\mathcal{S}_2,\mathcal{S}_1)/|\mathcal{S}_2|$, where $e(\mathcal{S}_1,\mathcal{S}_2)=\sum_{i\in\mathcal{S}_1,j\in\mathcal{S}_2}w_{ij}$ and $e(\mathcal{S}_2,\mathcal{S}_1)=\sum_{i\in\mathcal{S}_2,j\in\mathcal{S}_1}w_{ij}$, $|\mathcal{S}_i|$ denotes the cardinality of $\mathcal{S}_i$ for $i=1,2$.
\end{lemma}

\begin{lemma}[cf. \cite{MatrixAnalysis}]\label{Brauer's Theorem}
	Let $\Mb$ be an $n\times n$ arbitrary matrix with eigenvalues $\lambda_1,\lambda_2,\ldots,\lambda_n.$ Let $v$ be a right eigenvector of $\Mb$ associated with the eigenvalue $\lambda_k$, i.e., $\Mb v=\lambda_k v$, and let $q$ be any $n$-dimensional vector. Then the matrix $\Mb+vq^{\top}$ has eigenvalues $\lambda_1,\lambda_2,\ldots,\lambda_{k-1},\lambda_k+v^{\top}q,\lambda_{k+1},\ldots,\lambda_n$.
\end{lemma}

The following lemma shows  the continuous dependence of $\Phi^b(t,0)$ on $\hat{\Lb}^b(t)$, where $\Phi^b(t,0)$ is the state transition matrix of system \eqref{error-dynamics} corresponding to $\hat{\Lb}^b(t)$ defined on $[0,T]$. 
\smallskip
\begin{lemma}\label{state-transition-matrix-depend-on-laplacian-matrix}
Consider the linear interconnected system \eqref{linear-system-dynamics} communicating over $\mathcal{G}(t)$. Given any feedback matrix $\Kb$ and under Assumption \ref{boundedness}, $\Phi^b(t,0)$ depends continuously on $\hat{\Lb}^b(t)$ in the sense that for any $\epsilon>0$, there exists a $\delta>0$ such that $$\|\Phi^b_1(t,0)-\Phi^b_2(t,0)\|_{L^1(0,T)}\leq \epsilon$$ whenever $$\|\hat{\Lb}^b_1(t)-\hat{\Lb}^b_2(t)\|_{L^1(0,T)}\leq \delta,$$ where $\Phi^b_1(t,0)$ and $\Phi^b_2(t,0)$ are state transition matrices of system \eqref{error-dynamics} corresponding to $\hat{\Lb}^b_1(t)$ and $\hat{\Lb}^b_2(t)$, respectively.
\end{lemma}

We have been unable to find a proof of this intuitively reasonable result. Hence, we provide a proof in what follows.
\begin{IEEEproof}[Proof of Lemma \ref{state-transition-matrix-depend-on-laplacian-matrix}] 
Consider a convergent sequence $\{\hat{\Lb}^b_k(t) \}_{k=1}^\infty$ such that $$\|\hat{\Lb}^b_k(t)-\hat{\Lb}^b_*(t)\|_{L^1(0,T)}\to 0,\;\;  k\to \infty.$$ It suffices to show that $\|\Phi^b_k(t,0)\|$ converges to $\|\Phi^b_*(t,0)\|$, which are associated to $\hat{\Lb}_k^b(t,t_0)$ and $\hat{\Lb}^b_*(t)$, respectively,  in $L^1(0,T)$ as $k$ tends to infinity. Consider the following two linear systems governed, respectively, by:
\begin{numcases}{}
\dot{\yb}(t)=\left[\Ib\otimes \Ab-\hat{\Lb}^b_{k}(t)\otimes \Bb\Kb\right]\yb(t) \label{eqn-1}\\
\dot{\zb}(t)=\left[\Ib\otimes \Ab-\hat{\Lb}^b_{*}(t)\otimes \Bb\Kb\right]\zb(t) \label{eqn-2}
\end{numcases}
where $0\leq t\leq T$ and $\yb(0)=\zb(0)$.
Here, by Assumptions \ref{boundedness} and \ref{compactness-edge-weight}, $\hat{\Lb}^b_{k}(t)$ and $\hat{\Lb}^b_{*}(t)$ are bounded, so are $\yb(t)$ and $\zb(t)$ on $[0,T]$.
Let $\rb(t)=\yb(t)-\zb(t)$. $\rb(t)$ is also  bounded. The evolution of $\rb(t)$ is described by
\begin{align}\label{eqn-error-evolution}
\dot{\rb}=(\Ib\otimes \Ab)\rb-(\hat{\Lb}^b_{k}(t)\otimes \Bb\Kb)\rb+(\Delta \Lb(t)\otimes \Bb\Kb)\zb
\end{align}
where $\Delta \Lb(t)=\hat{\Lb}^b_{*}(t)-\hat{\Lb}^b_{k}(t)$. Denote by $\tilde{\Phi}(t,0)$ the state transition matrix of system \eqref{eqn-error-evolution}. Again, according to Assumption \ref{boundedness}, $\|\tilde{\Phi}(t,0)\|$ is bounded. Then,
\begin{align*}
\rb(t)=\tilde{\Phi}(t,0)\rb_0+\int_{t_0}^t \tilde{\Phi}(t,\tau)(\Delta \Lb(\tau)\otimes \Bb\Kb)\zb(\tau)\mathrm{d}\tau.
\end{align*}
Note that $\rb_0=\rb(0)=\zeb$.
Hence,
\begin{align*}
\|\rb(t)\|
\leq \int_{0}^t \big\|\tilde{\Phi}(t,\tau)\big\|\,\big\|\Delta \Lb(\tau)\otimes \Bb\Kb\big\|\,\big\|\zb(\tau)\big\|\mathrm{d}\tau. 
\end{align*}
Note that $\|\tilde{\Phi}(t,\tau)\|$ and $\left\|\zb(\tau)\right\|$ are bounded while given any $\epsilon>0$, there exists  a $K>0$ such that if $k>K$, then $\|\Delta \Lb(t)\|_{L^1(0,T)}<\epsilon$  by the convergence of $\{\hat{\Lb}^b_{k}(t)\}_{k=1}^{\infty}$  to $\hat{\Lb}^b_{*}(t)$ in $L^1(0,T)$. Then, one has that $\|\rb(t)\|_{L^1(0,T)}<M\epsilon$ if $k>K$ where $$M=T\sup_{\tau\in[0,T]}\|\tilde{\Phi}(t,\tau)\|\times\|\Bb\Kb\|\times\|\zb(\tau)\|<\infty.$$
Write
$\rb(t)=(\Phi^b_{m}(t,0)-\Phi_{*}^b(t,0))\yb(0)$, where $\Phi^b_{m}(t,0)$ and $\Phi_{*}^b(t,0)$ are state transition matrices of \eqref{eqn-1} and \eqref{eqn-2}, respectively.  Let $\yb(0)$ be an arbitrary unit vector, then if $k>K$, there holds  that $\|\Phi_{k}^b(t,0)-\Phi^b_{*}(t,0)\|_{L^1(0,T)}<M \epsilon$. This completes the proof. 
\end{IEEEproof}

We are now ready to prove Theorem \ref{necessity-connectivity-neutralstability}.  

Suppose  that $\mathcal{G}(t)$ is not jointly $(\delta,T)$-connected.
By the definition of joint $(\delta,T)$-connectivity, given any $T>0$ and any $\delta>0$, there exists a $t\geq 0$ such that $\bar{\mathcal{G}}=(\mathcal{V},\bar{\mathcal{E}},\bar{\Ab}=[\bar{w}_{ij}])$ is not connected, where $\bar{\mathcal{E}}$ contains edges satisfying $$(j,i):\;\bar{w}_{ij}=\int_t^{t+T}w_{ij}(s)\mathrm{d}s\geq \delta. $$  
As a result, there exists  a $\mathcal{S}_t\subset\mathcal{V}$ such that $\bar{w}_{pq}<\delta$ for $p\in\mathcal{S}_t$  and $q\in\mathcal{V}\setminus \mathcal{S}_t$. Hence,
$$
\sum_{i\in\mathcal{S}_t,\,j\in \mathcal{V}\setminus\mathcal{S}_t}\int_{t}^{t+T}w_{ij}(\tau)\mathrm{d}\tau\leq \delta\sum_{i\in\mathcal{S}_t,\,j\in \mathcal{V}\setminus\mathcal{S}_t}1\leq N^2\delta.
$$

Now fix $T>0$. Since $\delta>0$ is arbitrary, for any  $k\in\mathbb{Z}_{+}$, there exists a $t_k$ such that 
$
\sum_{i\in\mathcal{S}_k,\,j\in \mathcal{V}\setminus\mathcal{S}_k}\int_{t_k}^{t_k+T}w_{ij}(\tau)\mathrm{d}\tau\leq 1/(2k),
$
where we use $\mathcal{S}_k$ instead of $\mathcal{S}_{t_k}$ for simplicity. Note that $\mathcal{G}(t)$ is undirected, implying $w_{ij}(t)=w_{ji}(t)$ for any $i,j$. Then, 
\begin{align*}
\sum_{j\in\mathcal{S}_k,\,i\in \mathcal{V}\setminus\mathcal{S}_k}&\int_{t_k}^{t_k+T}w_{ij}(\tau)\mathrm{d}\tau\\
=&\sum_{i\in\mathcal{S}_k,\,j\in \mathcal{V}\setminus\mathcal{S}_k}\int_{t_k}^{t_k+T}w_{ij}(\tau)\mathrm{d}\tau\leq \frac{1}{2k}.
\end{align*}
By Lemma \ref{eigenvalue-bound}, $\lambda_2(\int_{t_{k}}^{t_{k}+T}\Lb(t)\mathrm{d}t)\leq 1/k$.

For each $t_k$, let $\hat{\Lb}_k^b(s)=\hat{\Lb}(s+t_k),s\in[0,T]$. Therefore, we have a sequence of matrix-valued functions $\{\hat{\Lb}^b_k(t)\in\mathcal{CLO}(\Sigma)\}_{k=1}^{\infty}$. By Lemma \ref{Brauer's Theorem}, 
$$
\sigma\left(\int_{0}^{T}\hat{\Lb}^b_k(t)\mathrm{d}t\right)=\left\{\lambda_i\left(\int_{0}^{T}\Lb^b_k(t)\mathrm{d}t\right),i\neq 1\right\}\cup\{1 \},
$$ 
where ``$\sigma(\cdot)$'' denotes the spectrum of a matrix.
Therefore,
$$
\lambda_1\left(\int_{0}^{T}\hat{\Lb}^b_k(t)\mathrm{d}t\right)=\lambda_2\left(\int_{0}^{T}\Lb^b_k(t)\mathrm{d}t\right)\leq\frac{1}{k}.
$$
Invoking Assumptions \ref{boundedness} and \ref{compactness-edge-weight}, there exists a convergent subsequence $\{\hat{\Lb}^b_{n_k}(t)\}$ of $\{\hat{\Lb}^b_k(t)\}$ in $\mathcal{CLO}(\Sigma)$.  For each $n_k$, 
$$
\lambda_1\left(\int_{0}^{T}\hat{\Lb}^b_{n_k}(t)\mathrm{d}t\right)\leq \frac{1}{n_k}.
$$
The limit  of $\hat{\Lb}^b_{n_k}(t)$ is denoted by $\lim_{k\to\infty}\hat{\Lb}^b_{n_k}(t)=\hat{\Lb}_*^b(t)\in\mathcal{CLO}(\Sigma).$

To prove Theorem \ref{necessity-connectivity-neutralstability}, we next characterize the lower and upper bounds for $\Fb_1(t)$ and $\Fb_2(t)$, respectively, with the aid of $\hat{\Lb}_*^b(t)$. Refer to \eqref{F-function} for the definitions of $\Fb_1(t)$ and $\Fb_2(t)$.

\medskip
\noindent (1) \textit{\textbf{Characterization of  $\Fb_2(t)$.}}
\medskip

Clearly, the matrix $\int_{0}^{T}\hat{\Lb}_*^b(t)\mathrm{d}t$ has a zero eigenvalue by the convergence of $\hat{\Lb}^b_{n_k}(t)$ to $\hat{\Lb}_*^b(t)$ as $k$ tends to infinity and the continuous dependence of eigenvalues on entries of a matrix. 
Recall that $\Phi^b_{n_k}(t,0)$ is the state transition matrix of system \eqref{error-dynamics} corresponding to $\hat{\Lb}^b_{n_k}(t)$. According to Lemma \ref{state-transition-matrix-depend-on-laplacian-matrix}, $\lim_{k\to\infty}\Phi^b_{n_k}(t,0)=\Phi^b_*(t,0),$ where $\Phi^b_*(t,0)$ is the state transition matrix of system \eqref{error-dynamics} corresponding to  $\hat{\Lb}_*^b(t)$. As a consequence, it is easily obtained  that
\begin{align*}
\lim_{k\to\infty}\big\|&(\Phi^b_{n_k})^\top(t,0)\left(\hat{\Lb}^b_{n_k}(t)\otimes \Pb\Bb\Bb^\top\Pb\right)\Phi^b_{n_k}(t,0)  \\
&-
(\Phi_*^b)^\top(t,0)\left(\hat{\Lb}_*^b(t)\otimes \Pb\Bb\Bb^\top\Pb\right)\Phi_*^b(t,0)\big\|_{L^1}=0.
\end{align*}

For subsequent analysis, let
$$
\Gamma_k=\int_{0}^{T}(\Phi_{n_k}^b)^\top(\tau,0)\left[\hat{\Lb}_{n_k}^b(\tau)\otimes \Pb\Bb\Bb^\top \Pb \right]\Phi_{n_k}^b(\tau,0)\mathrm{d}\tau,
$$
and $\Gamma_*$ be its limit, i.e., $\lim_{k\to\infty}\Gamma_k=\Gamma_*$.
By the definition of the set $\Sigma$ (see Assumption \ref{compactness-edge-weight}), if $\eb^\top(0)\Gamma_k\eb(0)\leq \eb^\top(0)\epsilon\eb(0)$ for  $k\in \mathbb{Z}_{+}$ and some $\epsilon>0$, then $\eb^\top(0)\Fb_2(t_{n_k})\eb(0)\leq \eb^\top(0)\epsilon \eb(0)$ for    $t_{n_k}$, as desired. (See \eqref{F-function} for the definition of $\Fb_2(t)$.) As a consequence, we will characterize  $\Gamma_k$ in what follows.

Let $\vb$ be a nullvector of $\int_0^{T}\hat{\Lb}^b_*(t)\mathrm{d}t$. Hence, $\hat{\Lb}^b_*(t)\vb\equiv 0$.  Let $\eb(0)=\vb\otimes \wb\in\mathbb{R}^{nN}$ be a nonzero initial state, where $\wb$ is any  nonzero vector in $\mathbb{R}^N$. Note that 
$$
(\hat{\Lb}^b_*(t)\otimes \Bb\Bb^\top \Pb)(\Ib\otimes e^{\Ab t})\eb(0)\equiv 0.
$$ 
Then, with $\Lb_*^b(t)$, it follows from system \eqref{error-dynamics} that $\eb(t)=(\Ib\otimes e^{\Ab t})\eb(0)$ for all $t\in[0,T]$. Consequently, 
$
\Phi_*^b(t,t_0)\eb(0)=(\Ib_{N}\otimes e^{\Ab t})\eb(0)
$
for $t\in[0,T]$. This gives that
\begin{align*}
\eb^\top&(0)\Gamma_*\eb(0)\\
&=\eb^\top(0)\left[\int_{0}^{T}\left[\Lb_{*}^b(\tau)\otimes (e^{\Ab t})^\top\Pb\Bb\Bb^\top \Pb e^{\Ab t} \right]\mathrm{d}\tau\right]\eb(0)\\
 &\overset{\eb(0)
 	=\vb\otimes \wb}{=} 0.
\end{align*}
Since the integral of real-valued functions is a linear continuous  operator, given arbitrary $\epsilon>0$, there exists a $K_1>0$ such that 
$
\eb^\top(0)\Gamma_k \eb(0)\leq \epsilon,
$
for all $k\geq K$. This immediately gives that $\eb^\top(0)\Fb_2(t_{n_k})\eb(0)\leq \epsilon$ for all $k\geq K_1$.
 
\medskip
\noindent (2) \textit{\textbf{Characterization of  $\Fb_1(t)$.}}
\medskip

Recall that  with $\hat{\Lb}_*^b(t)$, it follows from system \eqref{error-dynamics} that $\eb(t)=(\Ib\otimes e^{\Ab t})\eb(0)$ for all $t\in[0,T]$, where $\eb(0)=\vb\otimes \wb\in\mathbb{R}^{nN}$ is nonzero with $\wb$ being any nonzero vector in $\mathbb{R}^N$. Also, $\Phi_*^b(t,0)\eb(0)=(\Ib_{N}\otimes e^{\Ab t})\eb(0)$. Therefore,
\begin{align*}
\eb^\top(0) &\left[\int_{0}^{T}\left[\Phi_*^b(\tau,0)\right]^\top\left[\Ib\otimes (\Ab^\top\Pb+\Pb\Ab)\right]\left[\Phi_*^b(\tau,0)\right]  \mathrm{d}\tau\right]\eb(0)\\
&\hspace{-0.8cm}\overset{\eb(0)=\vb\otimes \wb}{=} \eb^\top(0)\left[\int_{0}^{T}\Ib\otimes \left[e^{\Ab t}\right]^\top(\Ab^\top\Pb+\Pb\Ab)\left[e^{\Ab t}\right] \mathrm{d}\tau\right]\eb(0) \\
=&\vb^\top\vb\times  \wb^\top\left[\int_{0}^{T} \left[e^{\Ab t}\right]^\top(\Ab^\top\Pb+\Pb\Ab)\left[e^{\Ab t}\right] \mathrm{d}\tau\right]\wb.
\end{align*}

We claim that there exists a nonzero $\wb^*$ such that $(\wb^*)^\top[\int_{0}^{T} [e^{\Ab t}]^\top(\Ab^\top\Pb+\Pb\Ab)[e^{\Ab t}] \mathrm{d}\tau]\wb^*\geq 0$. Now, we use the proof by contradiction to prove this claim. Suppose otherwise that  $$\lambda_{\max}\left(\int_{0}^{T} [e^{\Ab t}]^\top(\Ab^\top\Pb+\Pb\Ab)[e^{\Ab t}] \mathrm{d}\tau\right)< 0.$$   Consider the linear system $\dot{\zb}(t)=\Ab\zb(t)$ and let $E=\zb^\top(t)\Pb\zb(t)$. Then, integrating the derivative of $E(t)$ along  $\dot{\zb}(t)=\Ab\zb(t)$ gives
\begin{align*}
E(T+t)-E(t)=
\zb^\top(t)\Mb\zb(t)
\leq \frac{\lambda_{\max}(\Mb)}{\lambda_{\max}(\Pb)}E(t),
\end{align*}
where
$$
\Mb=\int_{0}^{T} \left[e^{\Ab t}\right]^\top(\Ab^\top\Pb+\Pb\Ab)\left[e^{\Ab t}\right] \mathrm{d}t.
$$
This shows that $E(t+T)\leq [1+(\lambda_{\max}(\Mb)/\lambda_{\max}(\Pb))]E(t)$. Since $\lambda_{\max}(\Mb)<0$ by the hypothesis and $E(t)\geq 0$ for all $t$, one has $-1\leq \lambda_{\max}(\Mb)/\lambda_{\max}(\Pb)<0$. Then, $E(t)$ converges to zero exponentially fast, so does $\zb(t)$ for any initial value $\zb(0)$. This contradicts the fact that $\sigma(\Ab)$ lies in the closed right-half plane. Hence, the existence of $\wb^*$ is ensured.

Now, considering again the fact that the integral of real-valued functions is a linear continuous  operator, given arbitrary $\epsilon>0$, there exists a  $K_2$ such that $\eb^\top(0)\Fb_1(t_{n_k})\eb(0)\geq -\epsilon$ for all $k\geq K_2$, where $\eb(0)=\vb\otimes \wb^*$. (See \eqref{F-function} for the definition of $\Fb_1(t)$.) 

\medskip 

Finally, given any $\epsilon>0$, by choosing $K=\max\{K_1,K_2 \}$, if $k\geq K$ then 
\begin{align}
\eb^\top(0)\left[\Fb_1(t_{n_k})-2\Fb_2(t_{n_k})\right]\eb(0)\geq -3\epsilon,
\end{align}
for $\eb(0)=\vb\otimes \wb^*$. This implies that $\Fb_1(t)-2\Fb_2(t)\leq -\alpha \Ib$ for some $\alpha>0$  does not hold uniformly in $t$ -- a contradiction. This completes the proof.

\subsection{\textsc{Proof of Theorem \ref{unstable-linear-systems}}}

The proof of Theorem \ref{unstable-linear-systems} makes use of the following lemma to characterize a useful property of an observable matrix pair $(\Ab,\Cb)$. 
\begin{lemma}\label{observability-over-positive-measure-set}
If $(\Ab,\Cb)$ is observable, then given any set $\Omega\subset [0,T]$ with $T>0$ that has a positive Lebesgue measure, one has $\int_{\Omega} (e^{\Ab \mu})^\top \Cb^\top \Cb e^{\Ab \mu}\mathrm{d} \mu >\zeb$.
\end{lemma}
We need the following result to prove Lemma \ref{observability-over-positive-measure-set}.
\begin{lemma}[cf. \cite{Mityagin}]\label{integral_positive_measure_set}
	Let $A(x)$ be a real analytic function on (a connected open domain $U$ of) $\mathbb{R}^d$. If $A$ is not identically zero, then its zero set 
	$F(A):=\{x\in U: A(x)=0 \}$
	has a zero measure.
\end{lemma}
 
\begin{IEEEproof}[Proof of Lemma \ref{observability-over-positive-measure-set}]
Suppose that $\int_{\Omega} (e^{\Ab \mu})^\top \Cb^\top \Cb e^{\Ab \mu}\mathrm{d} \mu >\zeb$ does not hold. Consequently, there exists a nonzero $\xb_0$ such that $$\xb_0^\top\left(\int_{\Omega} (e^{\Ab \mu})^\top \Cb^\top \Cb e^{\Ab \mu}\mathrm{d} \mu \right)\xb_0=0.$$
Let $f(\mu)=\xb_0^\top (e^{\Ab \mu})^\top \Cb^\top \Cb e^{\Ab \mu}\xb_0$. This is an analytic and real-valued function of $\mu$. Moreover, $f(\mu)$ is not a zero function on $(0,T)$ since $(\Ab,\Cb)$ is observable. Otherwise,  $$\xb_0^\top\left(\int_{[0,T]} (e^{\Ab \mu})^\top \Cb^\top \Cb e^{\Ab \mu}\mathrm{d} \mu \right)\xb_0=0,$$
which contradicts the observability. By Lemma \ref{integral_positive_measure_set}, the set $\{\mu| f(\mu)=0,0\leq \mu\leq T \}$ has a zero measure. Hence, $f(\mu)>0$ if $\mu\in \Omega\setminus \{\mu| f(\mu)=0,0\leq \mu\leq T \}$, namely, $f(\mu)$ is positive almost everywhere in $\Omega$. Then, $\int_{\Omega} f(\mu)\mathrm{d} \mu=0$ requires that $\Omega$ is a measure zero set -- a contradiction. This finishes the proof. 
\end{IEEEproof}

\smallskip

Given a controllable matrix pair $(\Ab,\Bb)$, the following lemma constructs an observable matrix pair $(\Ab,\Bb^\top\Pb)$.
\begin{lemma}\label{are-observability}
Let $\Re(\lambda(\Ab))\geq0 $ for any eigenvalue $\lambda(\Ab)$ of $\Ab$ and $\Pb$ be the solution of the following Riccati equation
\begin{align}\label{temp-1}
\Ab^\top \Pb+\Pb\Ab-\Pb\Rb\Pb+\Qb=\zeb
\end{align}
where  $\Rb=\Bb\Bb^\top\geq \zeb$, $(\Ab,\Bb)$ is controllable, and $\Qb\geq\zeb$ satisfying that $(\Ab,\Qb^{1/2})$ is observable. Then, $(\Ab,\Bb^\top\Pb)$ is observable.
\end{lemma}
\begin{IEEEproof}
Suppose otherwise that $(\Ab,\Bb^\top\Pb)$ is not observable. Then, there exists a $W\subset \mathbb{R}^n$ such that, for all $ \xb\in W$, $\Bb^\top\Pb\xb=\zeb$ and $\Ab\xb\in W$.  This implies that $e^{\Ab t}\xb\in W$ and $\Bb^\top\Pb e^{\Ab t} \xb=\zeb$ for all $\xb\in W$ and any $t$. Consider the linear system $\dot{\xb}=\Ab\xb$. Take $E=\xb^\top\Pb\xb$ as its Lyapunov function, whose derivative is
\begin{align*}
2\xb^\top\big(&\Ab^\top\Pb+\Pb\Ab\big)\xb\\
&=2\xb^\top(0)\left[e^{\Ab t}\right]^\top(\Ab^\top\Pb+\Pb\Ab)e^{\Ab t}\xb(0)\\
&\overset{\eqref{temp-1}}{=}2\xb^\top(0)\left[e^{\Ab t}\right]^\top(\Pb\Rb\Pb-\Qb)e^{\Ab t}\xb(0)\\
&\hspace{-0.5cm}\overset{\Bb^\top\Pb e^{\Ab t}\xb(0)=0}{=}2\xb^\top(0)\left[e^{\Ab t}\right]^\top(-\Qb)e^{\Ab t}\xb(0)\leq 0.
\end{align*}
Since $(\Ab,\Qb^{1/2})$ is observable, given $T>0$, there holds that 
$
\int_{s}^{s+T}\left[e^{\Ab t}\right]^\top\Qb\,\, e^{\Ab t}\mathrm{d}t>\zeb
$
for any $s\geq 0$.
This implies that $E(t+T)-E(t)\leq -\alpha E(t)$ for some $\alpha>0$, which directly gives that $E(t)$ converges to zero exponentially fast. However, since  $\Re(\lambda(\Ab))\geq0 $ for any eigenvalue $\lambda(\Ab)$ of $\Ab$, it is impossible that $E(t)$ converges to zero -- a contradiction.   This finishes the proof. 
\end{IEEEproof}

\smallskip
We are now ready to prove Theorem \ref{unstable-linear-systems}.
\smallskip

According to Lemma \ref{convergence-lemma}, it suffices to show the existence of positive real numbers
$a$ and $T$ such that for all $  t\ge 0$,
\begin{align}\label{alpha-property}
  \int_{t}^{t+T} \alpha(\tau)\mathrm{d}\tau\geq a.
\end{align}

({\bf Part I.}) By Assumption \ref{compactness-edge-weight}, the set $\Sigma$ is precompact, so that its closure, denoted by $\mathcal{CLO}(\Sigma)$, is a compact set. Note that $\mathcal{G}(t)$ is jointly $(\delta,T)$ connected. We first show that if $\hat{\Lb}_*^b(t)\in \mathcal{CLO}(\Sigma)$, then the integral of $\hat{\Lb}_*^b(t)$ over $[0,T]$ yields a connected $\delta$-graph (refer to Definition \ref{joint-connectivity} for the definition).   Since $\hat{\Lb}_*^b(t)\in \mathcal{CLO}(\Sigma)$, there exists a  $\{\hat{\Lb}^b_k(t)\in\Sigma \}_{k=1}^{\infty}$ such that $\lim_{k\to\infty}\hat{\Lb}_k^b(t)=\hat{\Lb}^b_*(t)$.

Let $\int_{0}^{T}w^b_{k,ij}(t)\mathrm{d}t\geq \delta$ for all $k$ and $w^b_{k,ij}(t)\to w^*_{ij}(t)$ as $k\to \infty$ for some $i,j\in\mathcal{V}$. It suffices to show that $$\int_{0}^{T} w^*_{ij}(t)\mathrm{d}t\geq \delta.$$ The convergence of $w^b_{k,ij}(t)$ implies  that given $\epsilon>0$, there exists a  $K(\epsilon)$ such that if $k>K(\epsilon)$, then
$$
\int_{0}^{T}|w^b_{k,ij}(t)-w^*_{ij}(t)|\leq \epsilon,
$$
which in turn yields
\begin{align*}
\int_{0}^{T} w^b_{k,ij}(t)\mathrm{d}t\leq  \epsilon T+\int_{0}^{T}w^*_{ij}(t)\mathrm{d}t.
\end{align*}
It follows in a straightforward manner that for a sufficiently large $k$, $\int_{0}^{T} w^b_{k,ij}(t)\mathrm{d}t<\delta,$ a contradiction. Hence, $\int_{0}^{T}w^*_{ij}(t)\mathrm{d}t\geq \delta$. This finishes the proof of the first part.

\smallskip

({\bf Part II.}) Now to prove the lower bound of \eqref{alpha-property}, we first show  that, for all $ \hat{\Lb}^b(t)\in\mathcal{CLO}(\Sigma)$,
\begin{align} 
\int_{0}^{T}(\Phi^b)^{\top}(\tau,0)\left[\hat{\Lb}^b(\tau)\otimes \Pb\Bb\Bb^\top \Pb \right]\Phi^b(\tau,0)\mathrm{d}\tau>\zeb\label{positive-definiteness},
\end{align}
where $\Phi^b(\cdot,\cdot)$ is the state transition matrix of system \eqref{error-dynamics} corresponding to $\hat{\Lb}^b(t)$.
In order to obtain a contradiction, suppose otherwise that there exists a  $\eb(0)$ such that
\begin{align}\label{tem-eqn-1}
\int_{0}^{T}\eb^\top(\tau)\left[\hat{\Lb}^b(\tau)\otimes \Pb\Bb\Bb^\top \Pb \right]\eb(\tau)\mathrm{d}\tau=0,
\end{align}
where $\eb(\tau)=\Phi^b(\tau,0)\eb(0)$ and $\hat{\Lb}^b(t)\in\mathcal{CLO}(\Sigma)$.
This gives that 
$$
\left(\hat{\Lb}^b(t)\otimes \Bb^\top \Pb\right)\eb(t)=\zeb,~~~~\forall t\in[0,T].
$$ 
Therefore, via \eqref{error-dynamics}, one has  $\eb(t)=e^{\Ab t}\eb(0)$ for $t\in[0,T]$, inserting which into \eqref{tem-eqn-1} in turn yields
\begin{align}\label{tem-eqn-2}
\int_{0}^{T}\eb^\top_{0}\left[\hat{\Lb}^b(\tau)\otimes (e^{\Ab\tau})^\top\Pb\Bb\Bb^\top \Pb e^{\Ab\tau} \right]\eb_0\,\mathrm{d}\tau=0.
\end{align}
 
We have assumed that $\mathcal{G}(t)$ is jointly $(\delta,T)$-connected. By the discussion in \textbf{Part I}, the union of the graph induced from $\hat{\Lb}^b(t)$ contains a connected $\delta$-graph. This implies that $\int_0^{T}\hat{\Lb}^b(t)\mathrm{d}t>\zeb$, which yields that there exists a set $\Omega\subset[0,T]$ such that $\eb^\top_{0}\left[\hat{\Lb}^b(\tau)\otimes \Ib \right]\eb_0>0$ for $\tau\in\Omega$ and $\Omega$ has a positive lebesgue measure. Hence, by \eqref{tem-eqn-2}, there must hold that 
\begin{align}\label{temp-3}
\int_{\tau\in\Omega} \eb^\top_{0}\left[\Ib\otimes (e^{\Ab\tau})^\top\Pb\Bb\Bb^\top \Pb e^{\Ab\tau} \right]\eb_0\,\mathrm{d}\tau=0.
\end{align}
Note however that $(\Ab,\Bb^\top\Pb)$ is observable. By Lemma \ref{observability-over-positive-measure-set}, equality \eqref{temp-3} does not hold, a contradiction.  This completes the proof in the second part.

\smallskip

({\bf Part III.})  In this part, we prove the existence of a positive definite lower bound for \eqref{positive-definiteness}.  

Again, we proceed with a contradiction argument. Suppose otherwise that there exist $\lambda_k=1/k\to 0$  as $k\to\infty$ such that
\begin{align*}
\lambda_{\min}\left[\int_{0}^{T}(\Phi^b_k)^\top(\tau,0)\left[\hat{\Lb}^b_k(\tau)\otimes \Pb\Bb\Bb^\top \Pb \right]\Phi^b_k(\tau,0)\mathrm{d}\tau\right]  
\leq \lambda_k,
\end{align*}
where $\hat{\Lb}_k^b(t)\in\mathcal{CLO}(\Sigma)$.
By Assumption \ref{compactness-edge-weight}, there exists a subsequence $\{\hat{\Lb}^b_{n_k}(t) \}_{k=1}^{\infty}$ such that it converges to $\hat{\Lb}^b_*(t)\in\mathcal{CLO}(\Sigma)$  as $k\to \infty$. Moreover, with $\hat{\Lb}^b_*(t)$, one has
\begin{align*}
\lambda_{\min}\left[\int_{0}^{T}\Phi_*^\top(\tau,0)\left[\hat{\Lb}_*(\tau)\otimes \Pb\Bb\Bb^\top \Pb \right]\Phi_*(\tau,0)\mathrm{d}\tau\right]=0,
\end{align*}
according to the continuity of integral operator and the continuous dependence of $\Phi^b(t,0)$ on $\hat{\Lb}^b(t)$ (see Lemma \ref{state-transition-matrix-depend-on-laplacian-matrix}).
This fact contradicts \eqref{positive-definiteness}.   Therefore, there exists an $a>0$ such that 
\begin{align}\label{tem-1}
\int_{0}^{T}(\Phi^b)^{\top}(\tau,0)\left[\hat{\Lb}^b(\tau)\otimes\Pb\Bb\Bb^\top \Pb \right]\Phi^b(\tau,0)\mathrm{d}\tau\geq a\Ib
\end{align}
for all $\hat{\Lb}^b(t)\in\mathcal{CLO}(\Sigma)$.


\smallskip
 
({\bf Part IV.}) In this final part, we prove GUEC. By \eqref{tem-eqn-3}, integrating $\dot{V}(t)$ over the interval $[0,T]$ gives the following relation:
\begin{align*}
\int_{t}^{t+T}\dot{V}\mathrm{d}t=&V(\eb(T+t))-V(\eb(t))  \\
=&\eb^\top(t)\left[\Fb_1(t)
-2\Fb_2(t)\right]\eb(t).
\end{align*}
Since $(\Ab,\Bb)$ is controllable and $(\Ab,\Qb^{1/2})$ is observable, given any $\kappa_1>0$,  there exists a unique $\Pb>\zeb$ such that
$$
\Ab^\top \Pb+\Pb\Ab-\kappa_1 \Pb\Bb\Bb^\top \Pb+ \Qb =\zeb.
$$
By Assumption \ref{compactness-edge-weight}, Lemma \ref{state-transition-matrix-depend-on-laplacian-matrix}, and the linearity of integral operator, $\int_{t}^{t+T}\Phi^\top(\tau,t)\left[\Ib \otimes \Pb\Bb\Bb^\top\Pb  \right]\Phi(\tau,t)\mathrm{d}\tau$ is upper bounded. Hence, there exists a positive constant $\kappa_2>0$ independent of $\kappa_1$ such that
\begin{align*}
\int_{t}^{t+T}&\Phi^\top(\tau,t)\left[\hat{\Lb}(\tau)\otimes \Pb\Bb\Bb^\top \Pb\right]\Phi(\tau,t)\mathrm{d}\tau\\
& \geq \frac{\kappa_2}{2} \underbrace{\int_{t}^{t+T}\Phi^\top(\tau,t)\left[\Ib \otimes \Pb\Bb\Bb^\top\Pb  \right]\Phi(\tau,t)\mathrm{d}\tau}_{\Fb_4(t)}
\end{align*}
for all $t$.
Consequently, one has 
\begin{align*}
\Fb_1(t)-2\Fb_2(t)&     
\leq \int_{t}^{t+T}\Phi^\top(\tau,t) \bigg[\Ib_N\otimes\Big(\Ab^\top\Pb+\Pb\Ab\\
&-\kappa_2 \Pb\Bb\Bb^\top \Pb\Big)\bigg]\Phi(\tau,t)\mathrm{d}\tau. \notag
\end{align*}
Now invoking the Theorem hypothesis that $\kappa_2\geq \kappa_1$, it follows  that there exists a real number $\kappa_3>0$ such that $\Fb_1(t)-2\Fb_2(t)\leq -\kappa_3\Ib$. Via similar arguments, one also has
$$\Fb_3(t)=\int_{t}^{t+T}\Phi^\top(\tau,t)\left(\Ib\otimes \Pb\right)\Phi(\tau,t)\mathrm{d}\tau\leq \kappa_4 \Ib$$
for some $\kappa_4>0$. One then has, for all $t\geq 0,$ that
$
\int_t^{t+T}\alpha(\tau)\mathrm{d}\tau\geq \kappa_3/\kappa_4>0.
$
According to Lemma \ref{convergence-lemma}, $ V(t)\leq \gamma_3e^{-\gamma_4 t}V(0)$ for appropriate $\gamma_3$ and $\gamma_4$. This finishes the proof.  

\section{Numerical Examples\label{sec:examples}}
\begin{example}\label{exp-1}
	In this  example, we illustrate Theorem \ref{unstable-linear-systems} with $\Ab$ being neutrally stable. Consider a set of four   linear systems interacting with each other. Set $$\Ab=\begin{bmatrix}0&1&0\\-2&0&1\\0&1&0\end{bmatrix}\ \text{ and }\ \Bb=\begin{bmatrix}0\\1\\1\end{bmatrix}.$$ $(\Ab,\Bb)$ is controllable. 
	Choose the Laplacian matrices (which correspond to two communication graphs $\mathcal{G}_1$ and $\mathcal{G}_2$ in Fig.~\ref{exp-3}, respectively) as follows:
	$$
	\Lb_1=0.1\begin{bmatrix}1&-1&0&0\\-1&1&0&0\\0&0&0&0\\0&0&0&0\end{bmatrix},\,\Lb_2=0.1\begin{bmatrix}2&0&-1&-1\\0&0&0&0\\-1&0&1&0\\-1&0&0&1\end{bmatrix}.
	$$	
	For illustration, the initial values are randomly chosen from $[-50,50]\times [-50,50]$. The underlying graph switches every $1s$ and $a_{14}(t)=a_{13}(t)=0.1-a_{12}(t)$. Specifically, $\mathcal{G}(t)=\mathcal{G}_1,\,2 k< t\leq 2 k+1$ and $\mathcal{G}(t)=\mathcal{G}_2,\,2 k+1<t\leq  2(k+1)$ for nonnegative integers $k$. Hence, the underlying interaction topology is uniformly jointly connected \cite{SuTAC2012}.

	By calculation, a positive definite solution of $\Ab^\top\Pb+\Pb\Ab=\zeb$ is
		$$
		\Pb^*=\begin{bmatrix}11&0&-8\\0&1.5&0\\-8&0&6.5\end{bmatrix}
		$$
		To illustrate Theorem  \ref{unstable-linear-systems}, note that   matrix $\Pb^*$ turns out to be a  positive definite solution of the following Riccati equation  
		\begin{align} 
			\Ab^\top \Pb+\Pb\Ab-\kappa_1 \Pb\Bb\Bb^\top \Pb+\Qb=\zeb
		\end{align}
		where  $\Qb=\kappa_1\Pb^*\Bb\Bb^\top\Pb^*$ for any $\kappa_1>0$ and $(\Ab,\Bb^\top\Pb^*)$ is observable, which is easy to verify. Moreover, $\Pb^*$ does \textit{not} depend on $\kappa_1$. Using $\Kb=\Bb^\top\Pb^*$, a sufficiently small $\kappa_1$ guarantees that the synchronization index is greater than 1.

	It can be observed from Fig.~\ref{fig-1-1}, which plots the logarithmic function of $V$, that the exponential convergence can be guaranteed and the convergence rate is indicated by the slope of the plot. 

	\begin{figure}
		\centering\includegraphics[width=9cm]{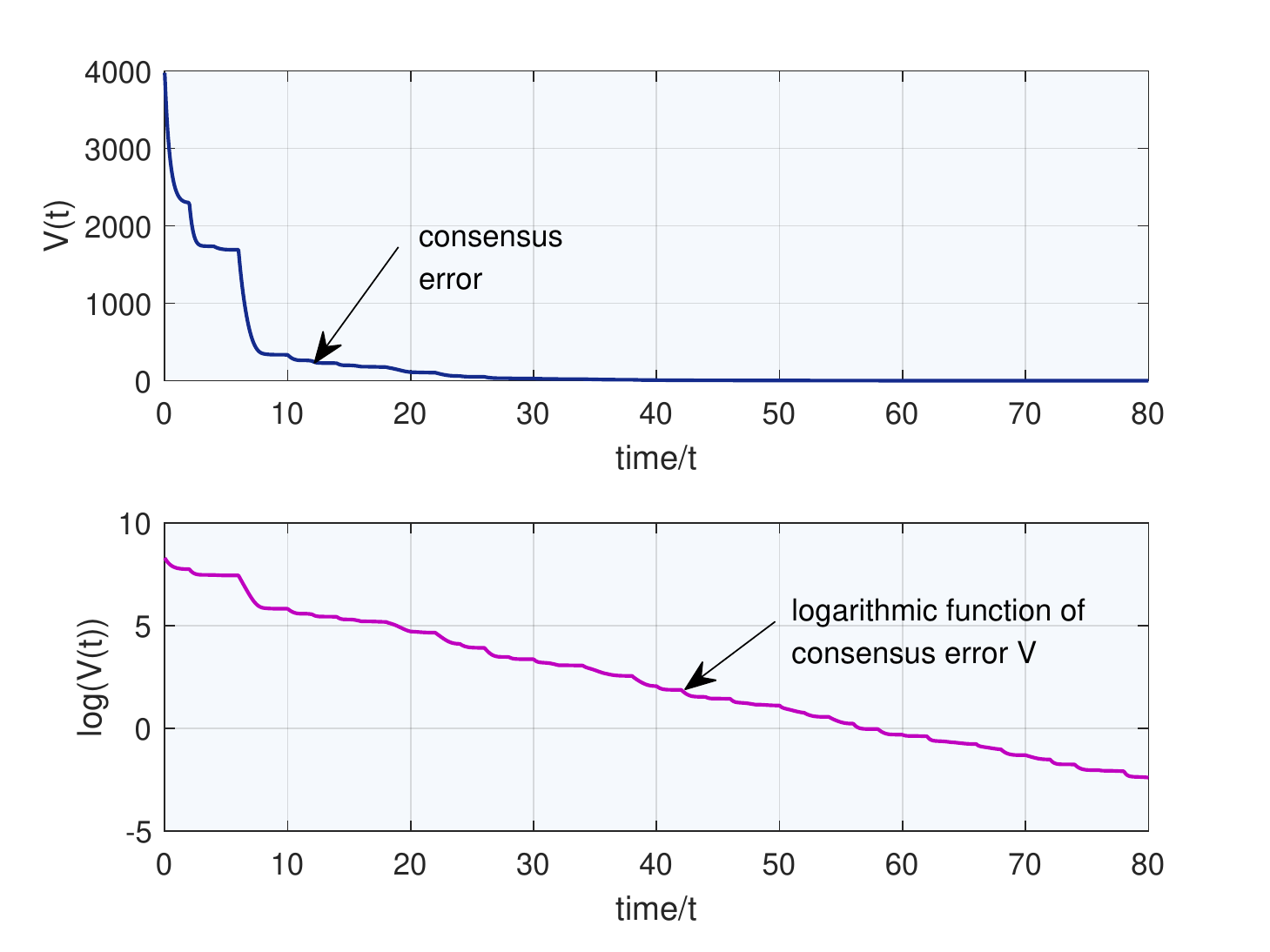}\caption{Trajectory of $\ln(V(t))$ when $G(t)$ is jointly connected (Example 1). }\label{fig-1-1}
	\end{figure}
	
\end{example}

\begin{example}\label{exp-2}
We illustrate Theorem \ref{unstable-linear-systems} with $\Ab$ having at least one unstable eigenvalue in this example. Let $\mathcal{G}(t)=\mathcal{G}_1,\,2 k< t\leq 2 k+1$ and $\mathcal{G}(t)=\mathcal{G}_2,\,2 k+1<t\leq  2(k+1)$ for nonnegative integers $k$, where $\mathcal{G}_1$ and $\mathcal{G}_2$ are shown in Fig.~\ref{exp-3}. The union of $\mathcal{G}(t)$ over any time interval of length $2$ is connected. Let
\begin{align*}
\Ab=\begin{bmatrix}
2&1&0\\
1&2&1\\
0&1&2\end{bmatrix},~
\Bb=\begin{bmatrix}
1&0\\
1&1\\
0&1\end{bmatrix}.
\end{align*}
	
Since $\mathcal{G}(t)$ is periodic, to perform Algorithm \ref{alg-1} and obtain $\kappa_2$, one only needs to calculate $\Fb_2(0)$ and $\Fb_4(0)$ with $T=2$.	Via simulation, the maximum and minimum eigenvalues of $\hat{\Pb}_{\gamma}=\Pb_{\gamma_k}/\hat{\lambda}(\Pb_{\gamma_k})$  are, respectively, plotted in Fig.~\ref{max_min_eigenvlaue}. ($\Pb_{\gamma_k}$ is scaled by dividing   its largest eigenvalue for ease of illustration.) Evidently, $\lambda_{\max}(\hat{\Pb}_{\gamma_k})=1$ while $\lambda_{\min}(\hat{\Pb}_{\gamma_k})$ converges to $0.072$ as $\gamma_k\to 0$. Fig.~\ref{feedbac_coefficient_computation} shows the minimum and the maximum eigenvalues of $\Fb_2(0)$ and $\Fb_4(0)$ with $T=2$, respectively. It is found that $\kappa_2$ can be chosen as $\kappa_2=0.042$, with which the matrix $\Pb$ is  
	$$
	\Pb=10^{3}*\begin{bmatrix}3.3367 &  -1.9075  &  2.3841\\
	-1.9075  &  1.9073 &  -1.907\\
	2.3841  & -1.9075  &  3.3367\end{bmatrix}.
	$$
	Fig.~\ref{error-evolution} plots the evolution of $\eb(t)$ with the above $\Pb$, which clearly converges to zero. The convergence rate is implied in Fig.~\ref{convergence_rate} via the slope of $\log(V(t))$.
	\begin{figure}
		\centering\includegraphics[width=8.5cm]{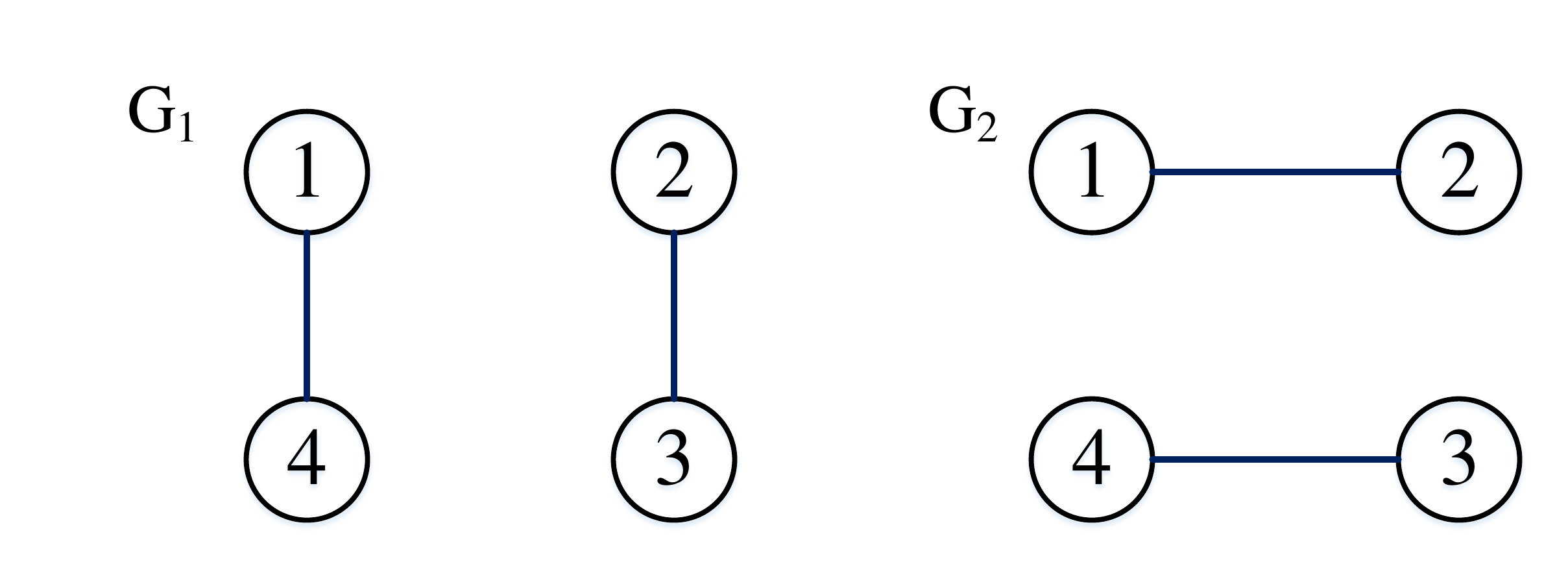}\caption{Two graphs over four nodes (Example 1). }\label{exp-3}
	\end{figure}
	
	\begin{figure}
		\centering\includegraphics[width=9cm]{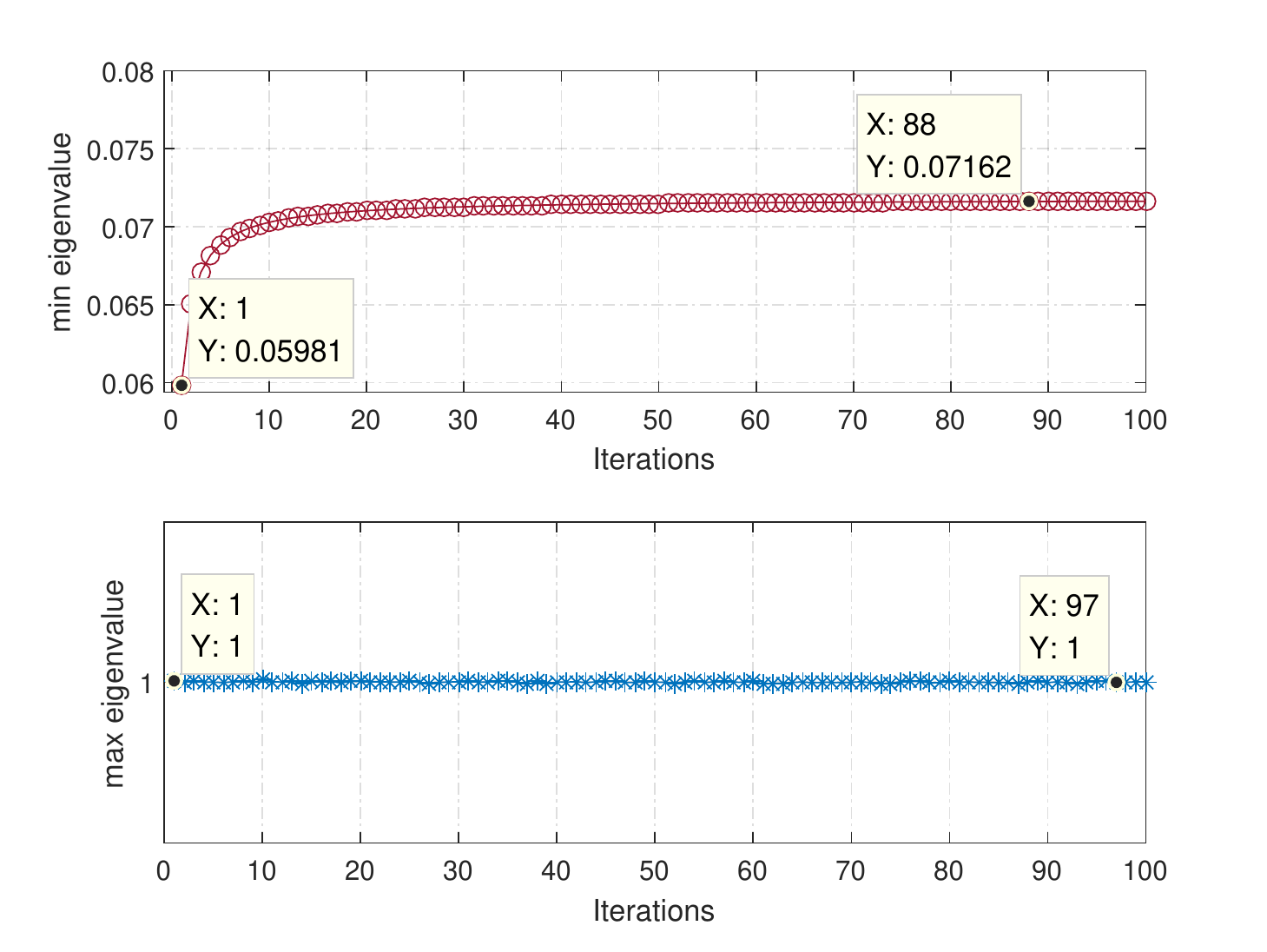}\caption{The minimum (the upper plot) and maximum eigenvalues (the lower plot) of $\hat{\Pb}=\Pb_{\gamma}/\lambda_{\max}(\Pb_{\gamma})$, respectively, with each $\gamma_k$ (Example 2).}\label{max_min_eigenvlaue}
	\end{figure}
	\begin{figure}
		\centering\includegraphics[width=9cm]{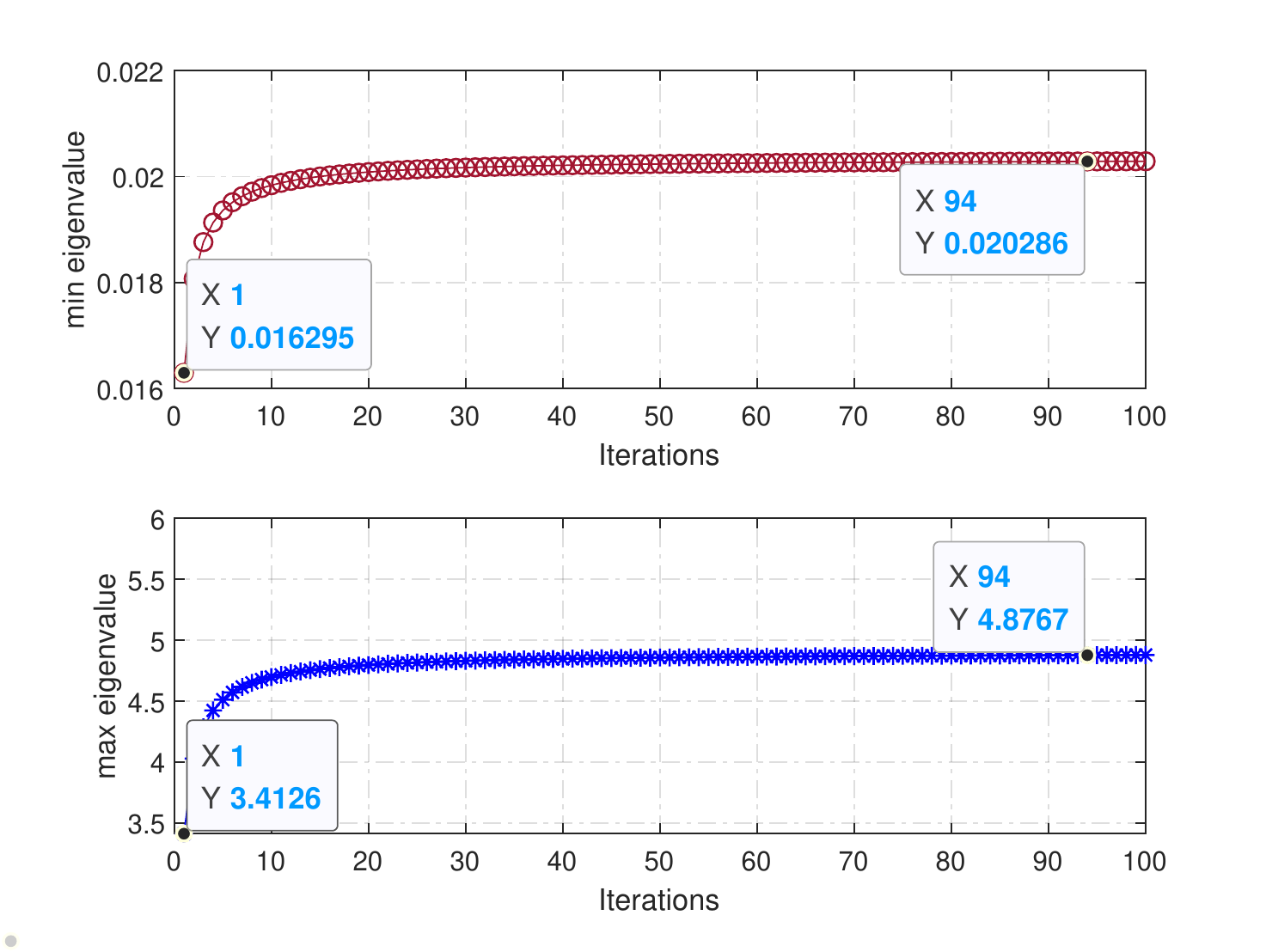}\caption{The minimum  eigenvalue of $\Fb_2(0)$ and the maximum eigenvalue of $\Fb_4(0)$ at each iteration (Example 2).}\label{feedbac_coefficient_computation}
	\end{figure}
	\begin{figure}
		\centering\includegraphics[width=9cm]{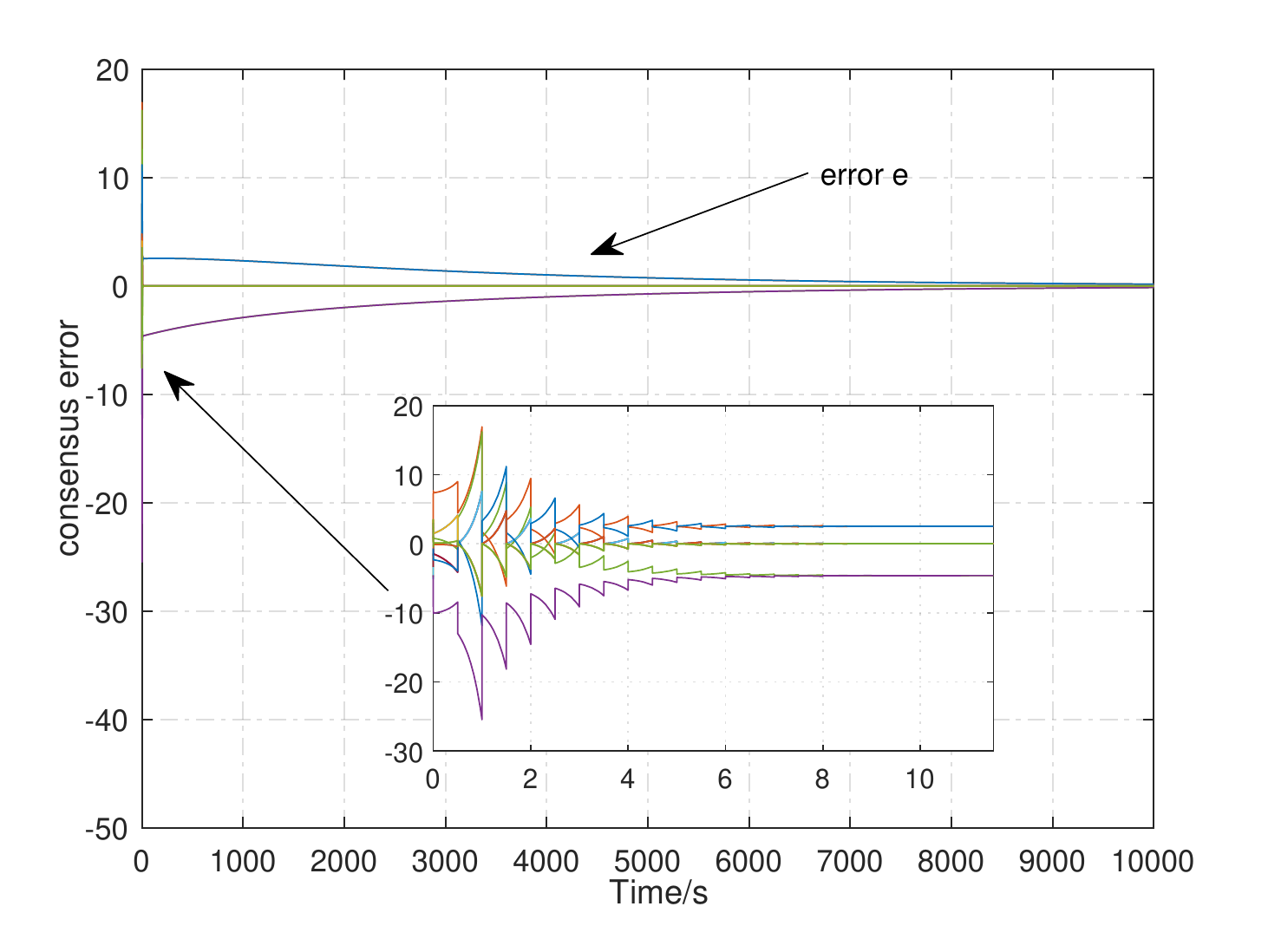}\caption{Evolution of consensus error $\eb(t)$ with a properly designed  feedback matrix (Example 2).}\label{error-evolution}
	\end{figure}
	\begin{figure}
		\centering\includegraphics[width=9cm]{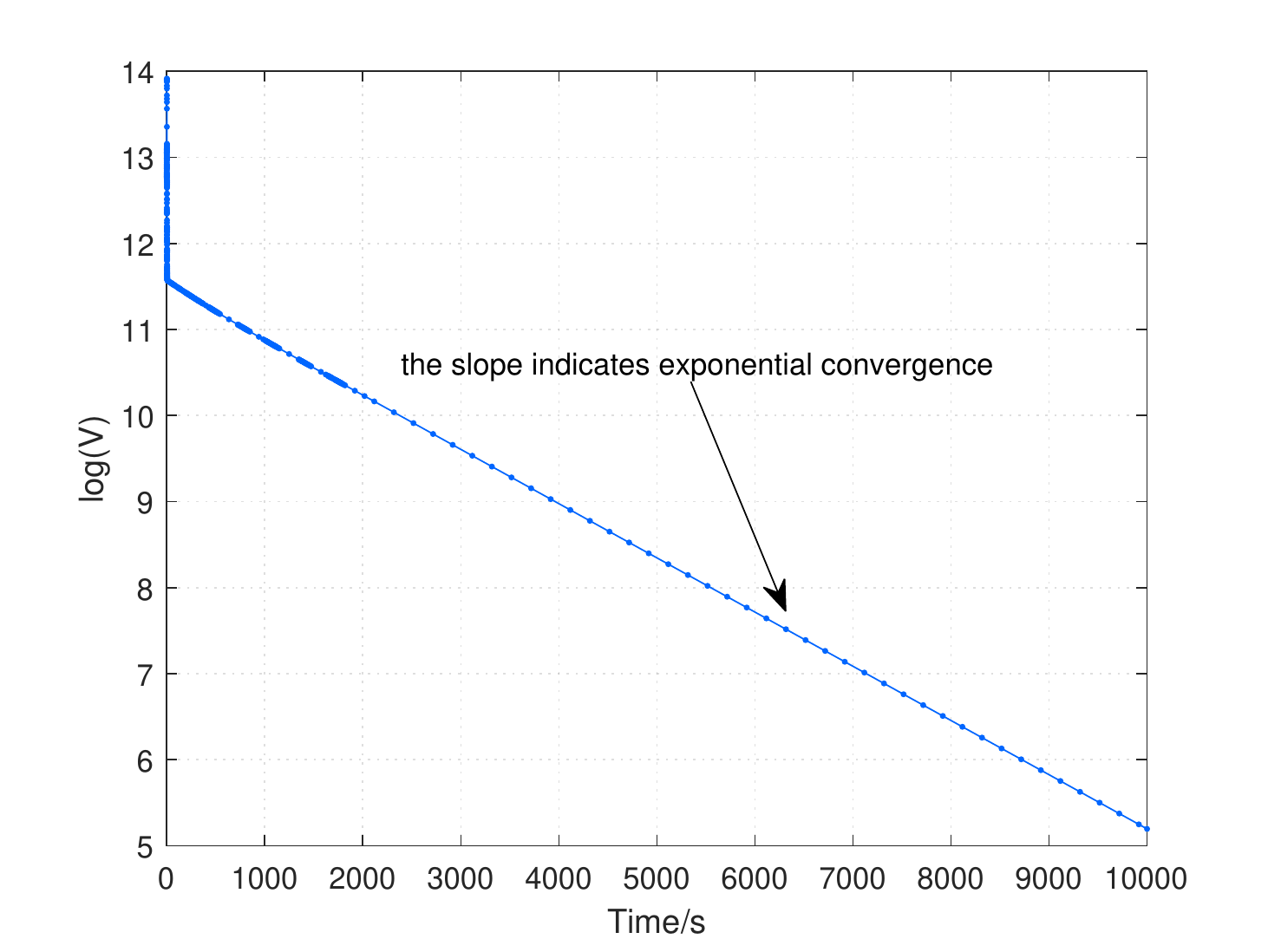}\caption{The trajectory of $\log(V(t))$ (Example 2).}\label{convergence_rate}
	\end{figure}
\end{example}

\begin{example}
In this example, we verify the necessity of controllability of $(\Ab,\Bb)$.	Consider a matrix pair $(\Ab,\Bb)$, which reads
	\begin{align*}
	\Ab=\begin{bmatrix}
	0&1&-1\\
	-1 &2 & -1\\
	0&-1&1 \end{bmatrix},~
	\Bb=\begin{bmatrix}
	1&0\\
	1  & 1\\
	1&2 \end{bmatrix}.
	\end{align*}
	$(\Ab,\Bb)$ is not controllable with $\lambda(\Ab)=3$ being the uncontrollable eigenvalue. Moreover, $\vb=[1,-2,1]^\top$ spans the uncontrollable space. Choose $$\Kb=\begin{bmatrix}1& 2& 1\\2& 1& 3\end{bmatrix},\;\;\Lb=\begin{bmatrix}1& -1\\-1 &1\end{bmatrix},$$ $\xb_1(0)=[1, -2, 1]^\top$ and $\xb_2(0)=[1, 1, 1]$. Consider the same time-varying graph $\mathcal{G}(t)$  as in Example \ref{exp-1}. The evolution  of $\vb^\top \xb_1(t)$ and $\vb^\top\xb_2(t)$ is shown in Fig.~\ref{projection}, which indicates that consensus cannot be reached. This is also verified in Fig.~\ref{state-evolution}, where the state trajectories clearly depict that no consensus is achieved.
	\begin{figure}
		\centering\includegraphics[width=9cm]{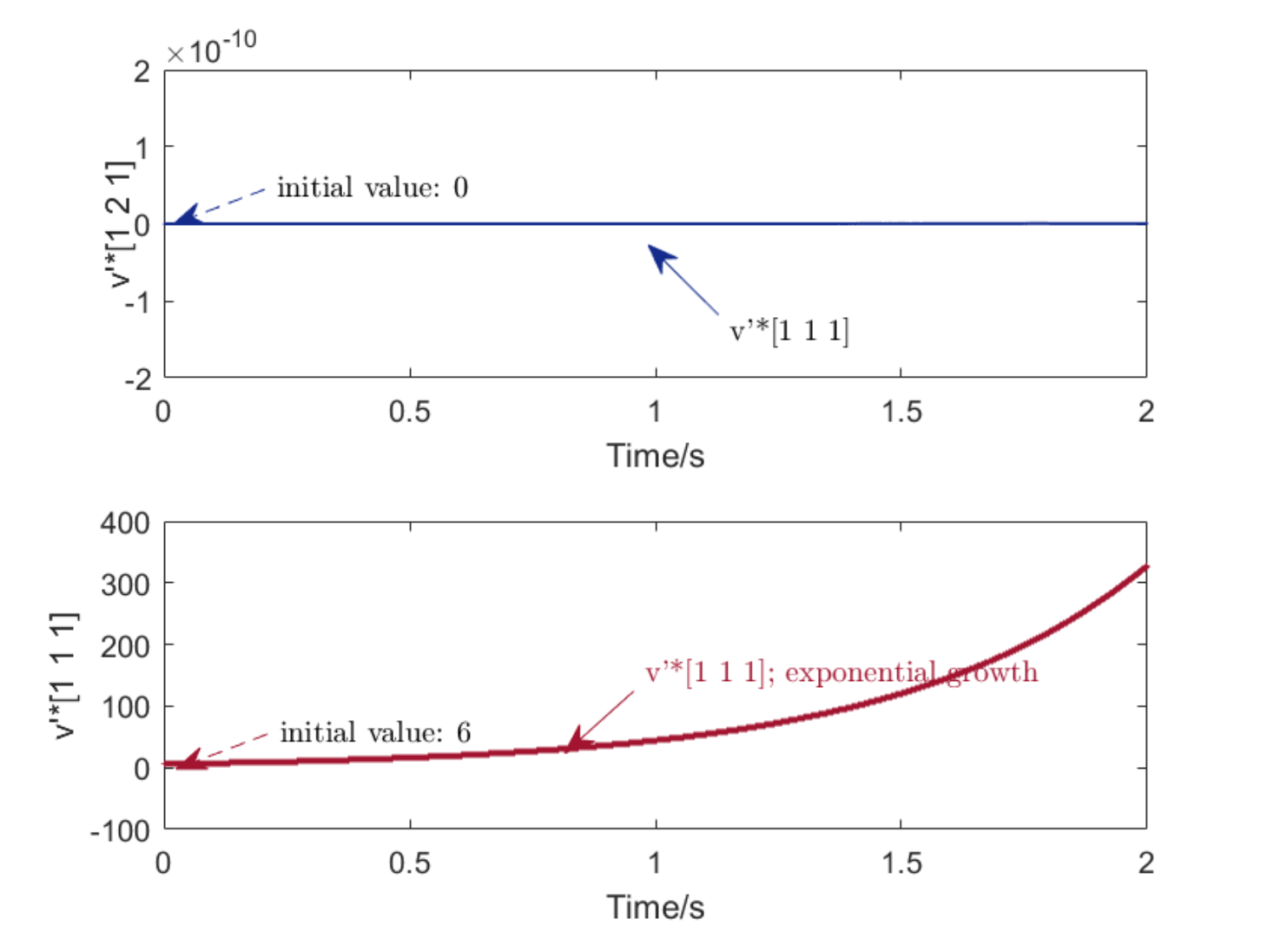}\caption{The evolution of $\vb^\top [1,-2,1]$ and $\vb^\top [1,1,1]$ (Example 3).}\label{projection}
	\end{figure}
	\begin{figure}
		\centering\includegraphics[width=9cm]{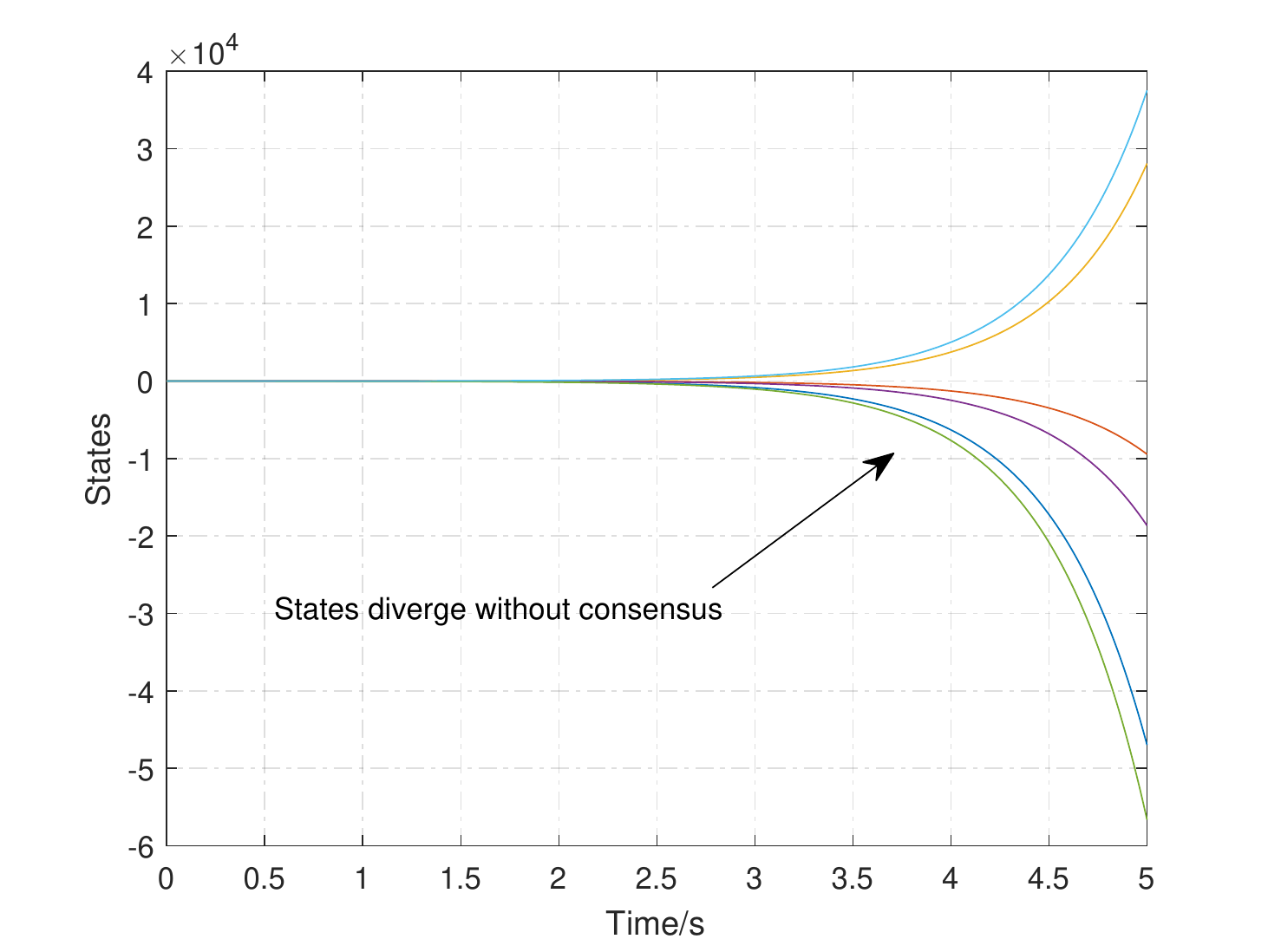}\caption{The evolution of states (Example 3).}\label{state-evolution}
	\end{figure}
\end{example}
\begin{example}\label{exp-delta-T-connectivity}
In this example, we illustrate the necessity of joint $(\delta,T)$-connectivity for exponential consensus. Let $\Ab=\zeb$ and $\Bb=\Kb=\Ib$. Let $\mathcal{G}_0,\, \mathcal{G}_1$ and $\mathcal{G}_2$ be undirected graphs with four nodes and edge weights equal to 1 (see Fig. \ref{exp2-fig1}).  
	\begin{figure}
		\centering\includegraphics[width=9cm]{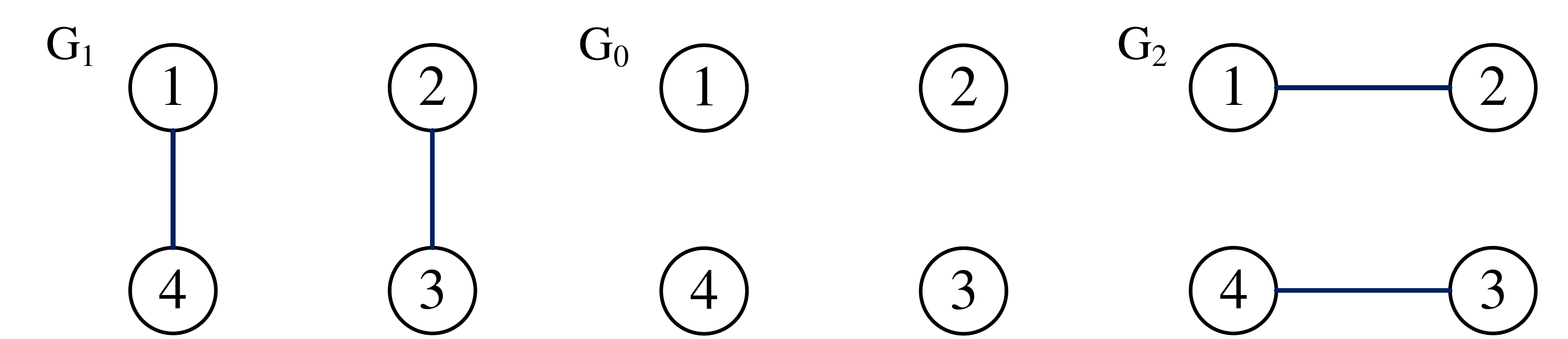}\caption{Three graphs which are jointly connected. None of the graphs is connected (Example 4).}\label{exp2-fig1}
	\end{figure}
	\begin{figure}
		\centering\includegraphics[width=8cm]{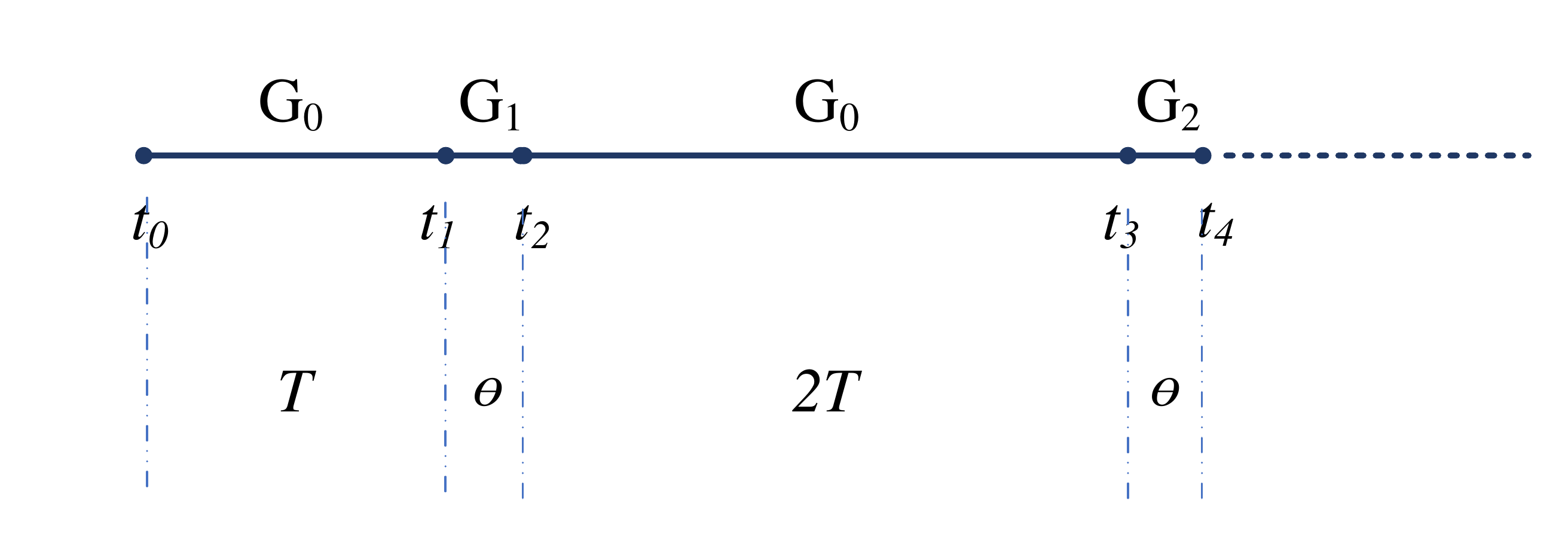}\caption{The switching scheme of the network topology among four nodes. $T$ and $\theta$ are both positive numbers (Example 4).}\label{exp2-fig2}
	\end{figure}
\begin{figure}
	\centering\includegraphics[width=9cm]{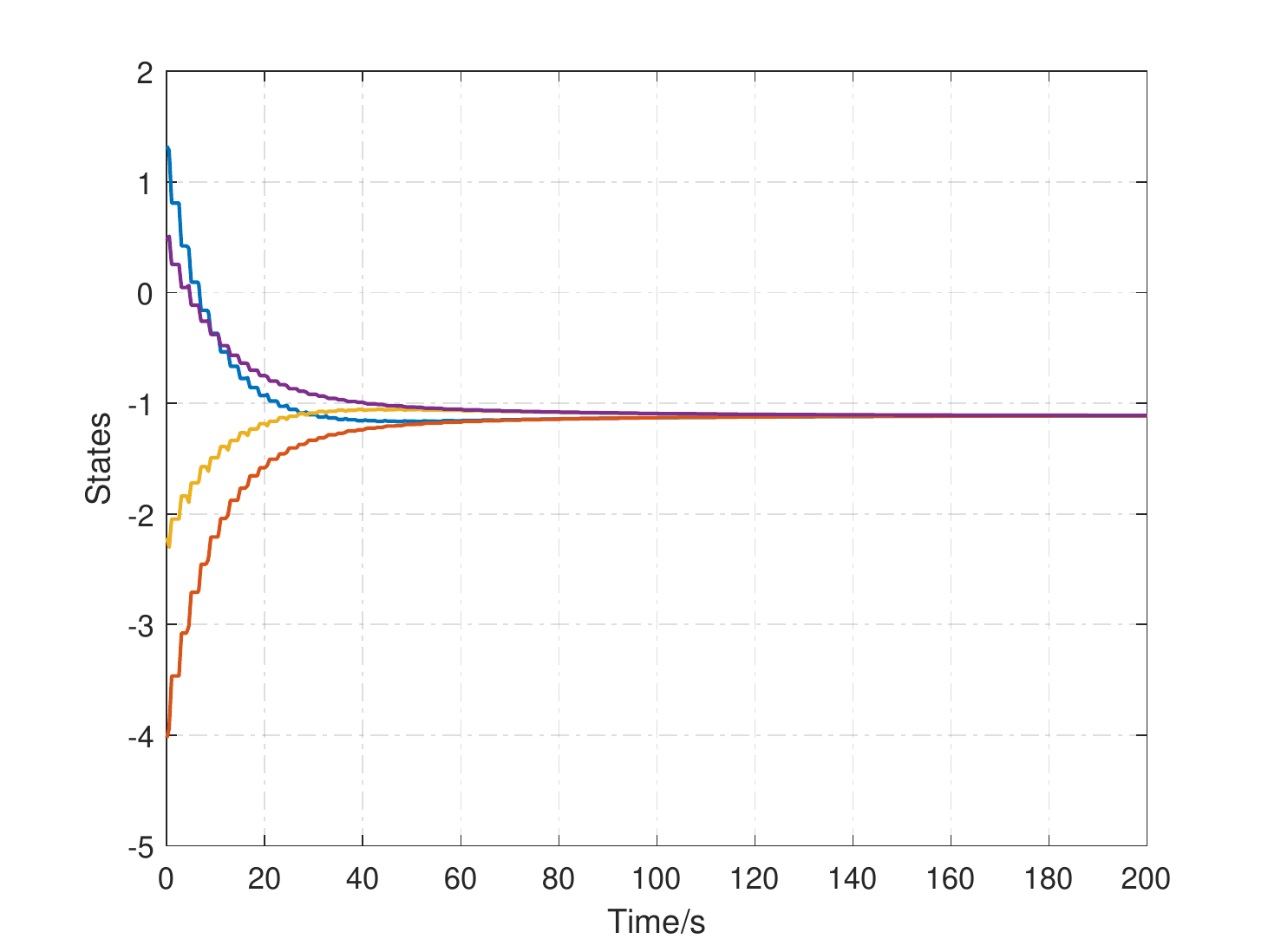}\caption{The evolution of states with switching scheme of the network topology shown in Fig.~\ref{exp2-fig2} (Example 4).}\label{exp4-consensus}
\end{figure}
	Consider the switching scheme shown in Fig.~\ref{exp2-fig2}. Specifically, $\mathcal{G}(t)=\mathcal{G}_{0}$ when $t\in[t_{2k},t_{2k}+(k+1)T),k=0,1,\ldots$ with $T=1$,  $\mathcal{G}(t)=\mathcal{G}_1,\;t\in[t_{4k+1},t_{4k+1}+\theta)$, and $\mathcal{G}(t)=\mathcal{G}_2,\;t\in[t_{4k+3},t_{4k+3}+\theta)$ where $\theta=0.5$. Note that $\mathcal{G}_0$ contains no edge.  Recall that $\alpha(t)$ is defined in \eqref{alpha-expression}. $$\int_{t_{2k}}^{t_{2k}+(k+1)T+\theta}\alpha(\tau)\mathrm{d}\tau\geq0.$$ Moreover, for any given $N>0$, there exists a $t_p$ such that for $t_{2k}>t_p$, $\int_{t_{2k}}^{t_{2k}+N}\alpha(\tau)\mathrm{d}\tau=0.$
	In this example, $V(t)$ converges to zero asymptotically as shown in Fig. \ref{exp4-consensus}, but not uniformly exponentially fast. To understand why the convergence is not exponential, suppose to the contrary that $$e^{c(t-s)+d}\geq e^{\int_{s}^{t}\alpha(\tau)\mathrm{d}\tau}\geq e^{a(t-s)+b}$$ for some real numbers $c\geq a>0$ and $b,d$.  This implies that there exists an $N>0$ such that $\int_{t_{2k}}^{t_{2k}+N}\alpha(\tau)\mathrm{d}\tau\geq a N+b>0$ for any $t_{2k}$, a contradiction.
\end{example}

\section{Conclusion\label{sec:conclusion}}
In this paper, we have investigated the global uniform exponential consensus problem for controllable linear systems  over time-varying undirected networks. A very mild joint connectivity condition on the communication graph has been proposed such that we can construct a set of matrix-valued functions that is precompact.   By designing a proper feedback matrix,
we have successfully shown that global uniform exponential consensus can be achieved  if the joint $(\delta,T)$-connectivity and controllability conditions  are satisfied, and in addition a synchronization index is greater than one. The necessity of  joint $(\delta,T)$-connectivity and controllability of linear systems for   global uniform exponential consensus have also been demonstrated with a properly designed quadratic Lyapunov function candidate. Finally, we point out that $\mathcal{G}(t)$ being undirected is somewhat restrictive. In view of this, the possible future works include the exploration of symmetric structure underlying  a directed switching network topology, by virtue of  matrix transformation \cite{QinTNNLS} or group theory \cite{LiberzonSCL}.


\begin{IEEEbiography}[{\includegraphics[width=1in,height=1.25in,clip]{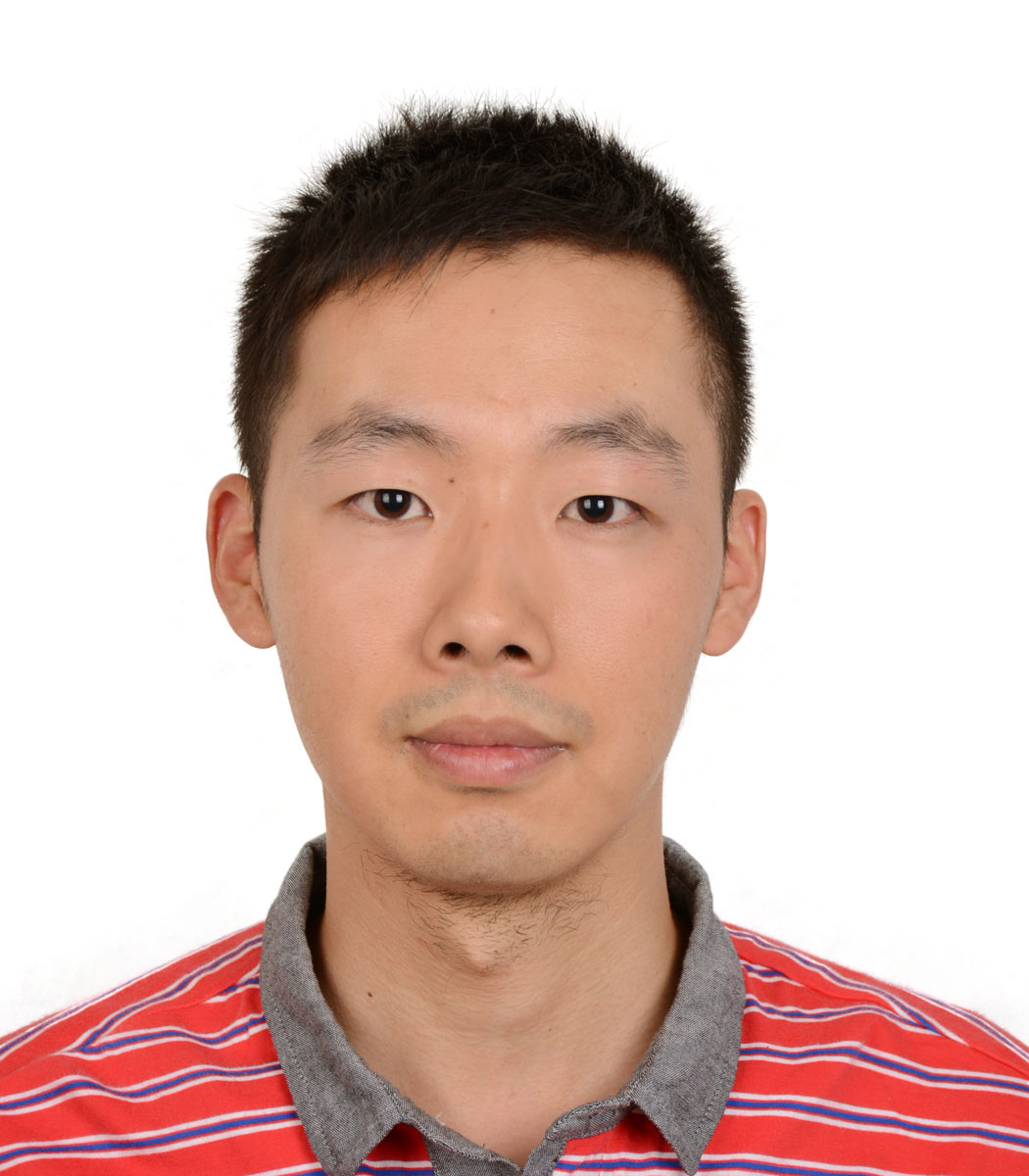}}]
	{Qichao Ma} received the Ph.D. degree in Control Science and Engineering from University of
	Science and Technology of China, Hefei, China,
	in 2019. From 2019 to 2021, he was a Postdoc Researcher with University of Science and
	Technology of China, Hefei, China, where he is
	currently a Research Associate Professor.
	
	His current research interests include distributed control, decision making, learning, and their applications.
\end{IEEEbiography}

\begin{IEEEbiography}[{\includegraphics[width=1in,height=1.25in,clip]{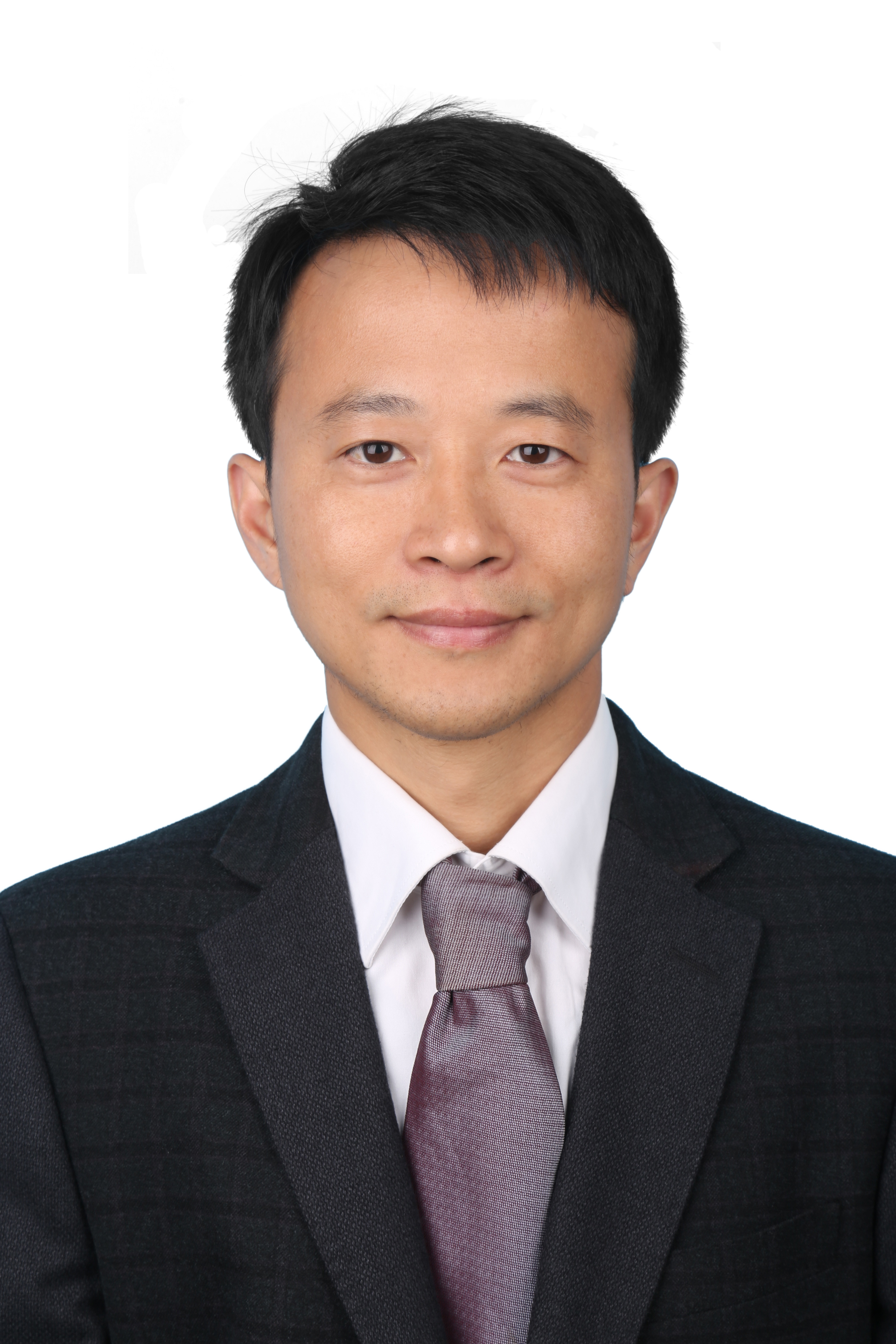}}]
	{Jiahu Qin}   received the first Ph.D. degree in control
	science and engineering from Harbin Institute of Technology, Harbin, China, in 2012, and the second Ph.D.
	degree in systems and control from The Australian
	National University, Canberra, ACT, Australia, in 2014.
	
	He is currently a Professor with the Department
	of Automation, University of Science and Technology
	of China, Hefei, China. His current research interests include networked control systems, autonomous
	intelligent systems, and human-robot interaction.
\end{IEEEbiography}
 

\begin{IEEEbiography}[{\includegraphics[width=1in,height=1.25in,clip]{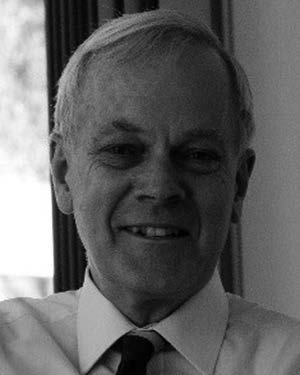}}]
	{Brian D. O. Anderson}  was born in Sydney, Australia, and educated at Sydney University in mathematics and electrical engineering, with PhD in electrical engineering from Stanford University in 1966.  
	
	 He was  appointed as ANU’s first engineering professor in 1981. He is now an Emeritus Professor at the Australian National University (having retired as Distinguished Professor in 2016). His awards include the IEEE Control Systems Award of 1997, the 2001 IEEE James H Mulligan, Jr Education Medal, and the Bode Prize of the IEEE Control System Society in 1992, as well as several IEEE and other best paper prizes. He is a Fellow of the Australian Academy of Science, the Australian Academy of Technological Sciences and Engineering, the Royal Society, and a foreign member of the US National Academy of Engineering. He holds honorary doctorates from a number of universities, including Université Catholique de Louvain, Belgium, and ETH, Zürich. He is a past president of the International Federation of Automatic Control and the Australian Academy of Science. 
	
\end{IEEEbiography}

 \vspace{-10cm}
  

\begin{IEEEbiography}[{\includegraphics[width=1in,height=1.25in,clip]{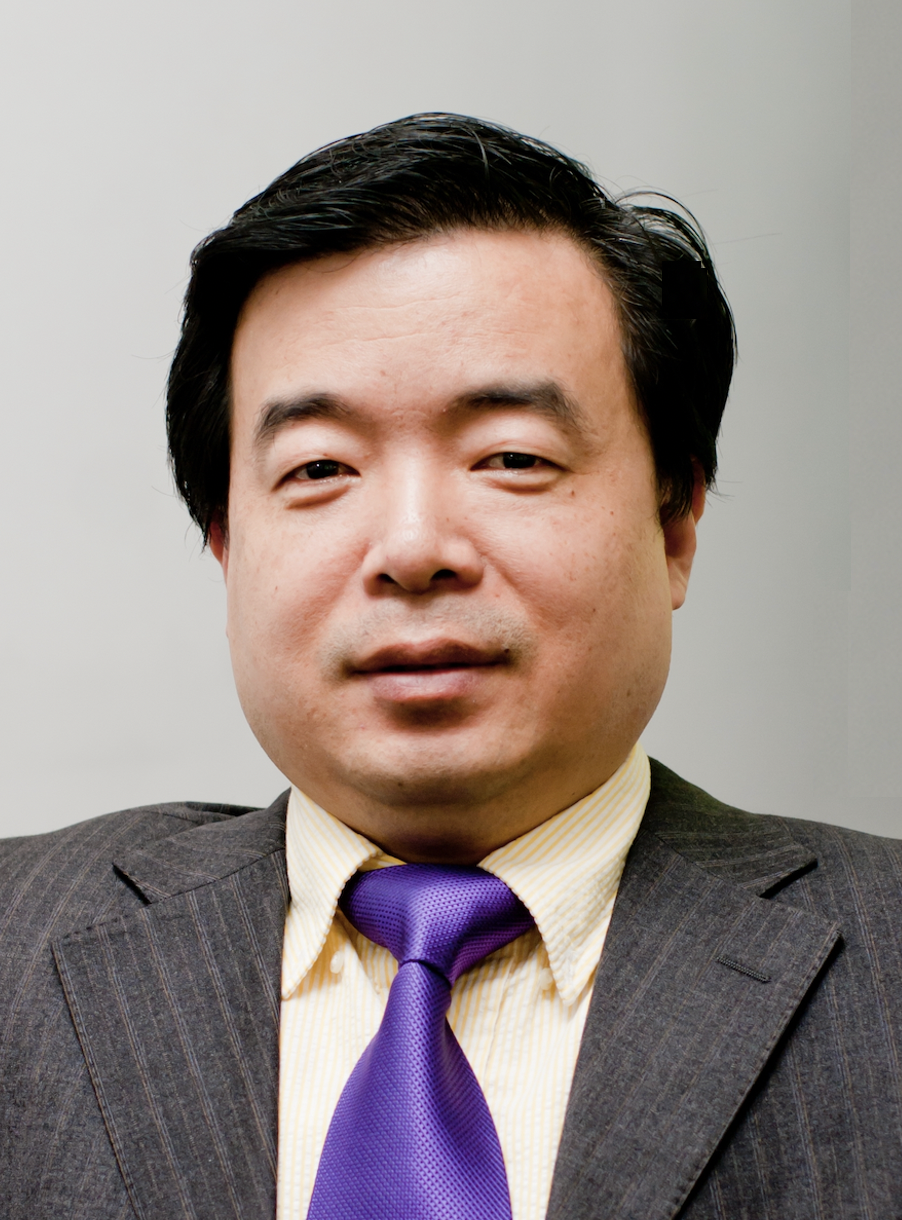}}]
	{Long Wang}  was born in Xi’an, China. He received the B.E. degree from Tsinghua University, Beijing, in 1986, and the Ph.D. degree from Peking University, Beijing, in 1992, both in dynamics and control.
	
	He has held research positions at the University of Toronto, Canada, and the German Aerospace Center, Munich, Germany. He is currently the Cheung Kong Chair Professor of Dynamics and Control, and the Director of Center for Systems and Control of Peking University. His research interests include complex networked systems, evolutionary game dynamics, artificial intelligence, and bio-mimetic robotics.
	
\end{IEEEbiography}

\end{document}